
%
%
%
\def\unredoffs{} 

%
%
%
%
\newbox\leftpage \newdimen\fullhsize \newdimen\hstitle \newdimen\hsbody
\tolerance=1000\hfuzz=2pt
\catcode`\@=11 
\magnification=1200\unredoffs\baselineskip=16pt plus 2pt minus 1pt
\hsbody=\hsize \hstitle=\hsize 
%
%
%
%
%
\newcount\yearltd\yearltd=\year\advance\yearltd by -1900

\def\Title#1#2{\nopagenumbers\abstractfont\hsize=\hstitle\rightline{#1}%
\vskip 1in\centerline{\titlefont #2}\abstractfont\vskip .5in\pageno=0}
\def\Date#1{\vfill\leftline{#1}\tenpoint\supereject\global\hsize=\hsbody%
\footline={\hss\tenrm\folio\hss}}
%

\def\draftmode{\message{ DRAFTMODE }\def\draftdate{{\rm preliminary draft:
\number\month/\number\day/\number\yearltd\ \ \hourmin}}%
\headline={\hfil\draftdate}\writelabels\baselineskip=20pt plus 2pt minus 2pt
 {\count255=\time\divide\count255 by 60 \xdef\hourmin{\number\count255}
  \multiply\count255 by-60\advance\count255 by\time
  \xdef\hourmin{\hourmin:\ifnum\count255<10 0\fi\the\count255}}}
\def\nolabels{\def\wrlabeL##1{}\def\eqlabeL##1{}\def\reflabeL##1{}}
\def\writelabels{\def\wrlabeL##1{\leavevmode\vadjust{\rlap{\smash%
{\line{{\escapechar=` \hfill\rlap{\sevenrm\hskip.03in\string##1}}}}}}}%
\def\eqlabeL##1{{\escapechar-1\rlap{\sevenrm\hskip.05in\string##1}}}%
\def\reflabeL##1{\noexpand\llap{\noexpand\sevenrm\string\string\string##1}}}
\nolabels
%
\global\newcount\secno \global\secno=0
\global\newcount\meqno \global\meqno=1
\def\newsec#1{\global\advance\secno by1\message{(\the\secno. #1)}
\global\subsecno=0\eqnres@t\noindent{\bf\the\secno. #1}
\writetoca{{\secsym} {#1}}\par\nobreak\medskip\nobreak}
\def\eqnres@t{\xdef\secsym{\the\secno.}\global\meqno=1\bigbreak\bigskip}
\def\sequentialequations{\def\eqnres@t{\bigbreak}}\xdef\secsym{}
\global\newcount\subsecno \global\subsecno=0
\def\subsec#1{\global\advance\subsecno by1\message{(\secsym\the\subsecno. #1)}
\ifnum\lastpenalty>9000\else\bigbreak\fi
\noindent{\it\secsym\the\subsecno. #1}\writetoca{\string\quad 
{\secsym\the\subsecno.} {#1}}\par\nobreak\medskip\nobreak}
\def\appendix#1#2{\global\meqno=1\global\subsecno=0\xdef\secsym{\hbox{#1.}}
\bigbreak\bigskip\noindent{\bf Appendix #1. #2}\message{(#1. #2)}
\writetoca{Appendix {#1.} {#2}}\par\nobreak\medskip\nobreak}
%
%
\def\eqnn#1{\xdef #1{(\secsym\the\meqno)}\writedef{#1\leftbracket#1}%
\global\advance\meqno by1\wrlabeL#1}
\def\eqna#1{\xdef #1##1{\hbox{$(\secsym\the\meqno##1)$}}
\writedef{#1\numbersign1\leftbracket#1{\numbersign1}}%
\global\advance\meqno by1\wrlabeL{#1$\{\}$}}
\def\eqn#1#2{\xdef #1{(\secsym\the\meqno)}\writedef{#1\leftbracket#1}%
\global\advance\meqno by1$$#2\eqno#1\eqlabeL#1$$}
%
\newskip\footskip\footskip14pt plus 1pt minus 1pt 
\def\footnotefont{\ninepoint}\def\f@t#1{\footnotefont #1\@foot}
\def\f@@t{\baselineskip\footskip\bgroup\footnotefont\aftergroup\@foot\let\next}
\setbox\strutbox=\hbox{\vrule height9.5pt depth4.5pt width0pt}
\global\newcount\ftno \global\ftno=0
\def\foot{\global\advance\ftno by1\footnote{$^{\the\ftno}$}}
%
\newwrite\ftfile   
\def\footend{\def\foot{\global\advance\ftno by1\chardef\wfile=\ftfile
$^{\the\ftno}$\ifnum\ftno=1\immediate\openout\ftfile=foots.tmp\fi%
\immediate\write\ftfile{\noexpand\smallskip%
\noexpand\item{f\the\ftno:\ }\pctsign}\findarg}%
\def\footatend{\vfill\eject\immediate\closeout\ftfile{\parindent=20pt
\centerline{\bf Footnotes}\nobreak\bigskip\input foots.tmp }}}
\def\footatend{}
%
%
\global\newcount\refno \global\refno=1
\newwrite\rfile
\def\ref{[\the\refno]\nref}
\def\nref#1{\xdef#1{[\the\refno]}\writedef{#1\leftbracket#1}%
\ifnum\refno=1\immediate\openout\rfile=refs.tmp\fi
\global\advance\refno by1\chardef\wfile=\rfile\immediate
\write\rfile{\noexpand\item{#1\ }\reflabeL{#1\hskip.31in}\pctsign}\findarg}
\def\findarg#1#{\begingroup\obeylines\newlinechar=`\^^M\pass@rg}
{\obeylines\gdef\pass@rg#1{\writ@line\relax #1^^M\hbox{}^^M}%
\gdef\writ@line#1^^M{\expandafter\toks0\expandafter{\striprel@x #1}%
\edef\next{\the\toks0}\ifx\next\em@rk\let\next=\endgroup\else\ifx\next\empty%
\else\immediate\write\wfile{\the\toks0}\fi\let\next=\writ@line\fi\next\relax}}
\def\striprel@x#1{} \def\em@rk{\hbox{}} 
\def\lref{\begingroup\obeylines\lr@f}
\def\lr@f#1#2{\gdef#1{\ref#1{#2}}\endgroup\unskip}

\def\addref#1{\immediate\write\rfile{\noexpand\item{}#1}} 
\def\startrefs#1{\immediate\openout\rfile=refs.tmp\refno=#1}
\def\xref{\expandafter\xr@f}\def\xr@f[#1]{#1}
\def\refs#1{\count255=1[\r@fs #1{\hbox{}}]}
\def\r@fs#1{\ifx\und@fined#1\message{reflabel \string#1 is undefined.}%
\nref#1{need to supply reference \string#1.}\fi%
\vphantom{\hphantom{#1}}\edef\next{#1}\ifx\next\em@rk\def\next{}%
\else\ifx\next#1\ifodd\count255\relax\xref#1\count255=0\fi%
\else#1\count255=1\fi\let\next=\r@fs\fi\next}
%

%
\newwrite\ffile\global\newcount\figno \global\figno=1
\def\fig{fig.~\the\figno\nfig}
\def\nfig#1{\xdef#1{fig.~\the\figno}%
\writedef{#1\leftbracket fig.\noexpand~\the\figno}%
\ifnum\figno=1\immediate\openout\ffile=figs.tmp\fi\chardef\wfile=\ffile%
\immediate\write\ffile{\noexpand\medskip\noexpand\item{Fig.\ \the\figno. }
\reflabeL{#1\hskip.55in}\pctsign}\global\advance\figno by1\findarg}
\def\xfig{\expandafter\xf@g}\def\xf@g fig.\penalty\@M\ {}
\def\figs#1{figs.~\f@gs #1{\hbox{}}}
\def\f@gs#1{\edef\next{#1}\ifx\next\em@rk\def\next{}\else
\ifx\next#1\xfig #1\else#1\fi\let\next=\f@gs\fi\next}
\newwrite\lfile
{\escapechar-1\xdef\pctsign{\string\%}\xdef\leftbracket{\string\{}
\xdef\rightbracket{\string\}}\xdef\numbersign{\string\#}}

\def\writestop{\def\writestoppt{\immediate\write\lfile{\string\pageno%
\the\pageno\string\startrefs\leftbracket\the\refno\rightbracket%
\string\def\string\secsym\leftbracket\secsym\rightbracket%
\string\secno\the\secno\string\meqno\the\meqno}\immediate\closeout\lfile}}
\def\writestoppt{}\def\writedef#1{}
\def\seclab#1{\xdef #1{\the\secno}\writedef{#1\leftbracket#1}\wrlabeL{#1=#1}}
\def\subseclab#1{\xdef #1{\secsym\the\subsecno}%
\writedef{#1\leftbracket#1}\wrlabeL{#1=#1}}
\newwrite\tfile \def\writetoca#1{}
\def\leaderfill{\leaders\hbox to 1em{\hss.\hss}\hfill}
\def\writetoc{\immediate\openout\tfile=toc.tmp 
   \def\writetoca##1{{\edef\next{\write\tfile{\noindent ##1 
   \string\leaderfill {\noexpand\number\pageno} \par}}\next}}}

%
\catcode`\@=12 
%
\edef\tfontsize{\ifx\answ\bigans scaled\magstep3\else scaled\magstep4\fi}
\font\titlerm=cmr10 \tfontsize \font\titlerms=cmr7 \tfontsize
\font\titlermss=cmr5 \tfontsize \font\titlei=cmmi10 \tfontsize
\font\titleis=cmmi7 \tfontsize \font\titleiss=cmmi5 \tfontsize
\font\titlesy=cmsy10 \tfontsize \font\titlesys=cmsy7 \tfontsize
\font\titlesyss=cmsy5 \tfontsize \font\titleit=cmti10 \tfontsize
\skewchar\titlei='177 \skewchar\titleis='177 \skewchar\titleiss='177
\skewchar\titlesy='60 \skewchar\titlesys='60 \skewchar\titlesyss='60
\def\titlefont{\def\rm{\fam0\titlerm}
\textfont0=\titlerm \scriptfont0=\titlerms \scriptscriptfont0=\titlermss
\textfont1=\titlei \scriptfont1=\titleis \scriptscriptfont1=\titleiss
\textfont2=\titlesy \scriptfont2=\titlesys \scriptscriptfont2=\titlesyss
\textfont\itfam=\titleit \def\it{\fam\itfam\titleit}\rm}
 \ifx\answ\bigans\else scaled\magstep1\fi
\ifx\answ\bigans\def\abstractfont{\tenpoint}\else
\font\abssl=cmsl10 scaled \magstep1
\font\absrm=cmr10 scaled\magstep1 \font\absrms=cmr7 scaled\magstep1
\font\absrmss=cmr5 scaled\magstep1 \font\absi=cmmi10 scaled\magstep1
\font\absis=cmmi7 scaled\magstep1 \font\absiss=cmmi5 scaled\magstep1
\font\abssy=cmsy10 scaled\magstep1 \font\abssys=cmsy7 scaled\magstep1
\font\abssyss=cmsy5 scaled\magstep1 \font\absbf=cmbx10 scaled\magstep1
\skewchar\absi='177 \skewchar\absis='177 \skewchar\absiss='177
\skewchar\abssy='60 \skewchar\abssys='60 \skewchar\abssyss='60
\def\abstractfont{\def\rm{\fam0\absrm}
\textfont0=\absrm \scriptfont0=\absrms \scriptscriptfont0=\absrmss
\textfont1=\absi \scriptfont1=\absis \scriptscriptfont1=\absiss
\textfont2=\abssy \scriptfont2=\abssys \scriptscriptfont2=\abssyss
\textfont\itfam=\bigit \def\it{\fam\itfam\bigit}\def\footnotefont{\tenpoint}%
\textfont\slfam=\abssl \def\sl{\fam\slfam\abssl}%
\textfont\bffam=\absbf \def\bf{\fam\bffam\absbf}\rm}\fi
\def\tenpoint{\def\rm{\fam0\tenrm}
\textfont0=\tenrm \scriptfont0=\sevenrm \scriptscriptfont0=\fiverm
\textfont1=\teni  \scriptfont1=\seveni  \scriptscriptfont1=\fivei
\textfont2=\tensy \scriptfont2=\sevensy \scriptscriptfont2=\fivesy
\textfont\itfam=\tenit \def\it{\fam\itfam\tenit}\def\footnotefont{\ninepoint}%
\textfont\bffam=\tenbf \def\bf{\fam\bffam\tenbf}\def\sl{\fam\slfam\tensl}\rm}
\font\ninerm=cmr9 \font\sixrm=cmr6 \font\ninei=cmmi9 \font\sixi=cmmi6 
\font\ninesy=cmsy9 \font\sixsy=cmsy6 \font\ninebf=cmbx9 
\font\nineit=cmti9 \font\ninesl=cmsl9 \skewchar\ninei='177
\skewchar\sixi='177 \skewchar\ninesy='60 \skewchar\sixsy='60 
\def\ninepoint{\def\rm{\fam0\ninerm}
\textfont0=\ninerm \scriptfont0=\sixrm \scriptscriptfont0=\fiverm
\textfont1=\ninei \scriptfont1=\sixi \scriptscriptfont1=\fivei
\textfont2=\ninesy \scriptfont2=\sixsy \scriptscriptfont2=\fivesy
\textfont\itfam=\ninei \def\it{\fam\itfam\nineit}\def\sl{\fam\slfam\ninesl}%
\textfont\bffam=\ninebf \def\bf{\fam\bffam\ninebf}\rm} 
%
%
\def\noblackbox{\overfullrule=0pt}
\hyphenation{anom-aly anom-alies coun-ter-term coun-ter-terms}
\def\inv{^{\raise.15ex\hbox{${\scriptscriptstyle -}$}\kern-.05em 1}}

\def\Dsl{\,\raise.15ex\hbox{/}\mkern-13.5mu D} 
\def\dsl{\raise.15ex\hbox{/}\kern-.57em\partial}

\font\bigit=cmti10 scaled \magstep1
\def\lspace{\ifx\answ\bigans{}\else\qquad\fi}
\def\lbspace{\ifx\answ\bigans{}\else\hskip-.2in\fi} 
\def\boxeqn#1{\vcenter{\vbox{\hrule\hbox{\vrule\kern3pt\vbox{\kern3pt
        \hbox{${\displaystyle #1}$}\kern3pt}\kern3pt\vrule}\hrule}}}
\def\mbox#1#2{\vcenter{\hrule \hbox{\vrule height#2in
                \kern#1in \vrule} \hrule}}  
%

\def\darr#1{\raise1.5ex\hbox{$\leftrightarrow$}\mkern-16.5mu #1}

\def\half{{\textstyle{1\over2}}} 
\def\roughly#1{\raise.3ex\hbox{$#1$\kern-.75em\lower1ex\hbox{$\sim$}}}

\openup -1pt
\input epsf
\expandafter\ifx\csname pre amssym.tex at\endcsname\relax \else\endinput\fi
\expandafter\chardef\csname pre amssym.tex at\endcsname=\the\catcode`\@
\catcode`\@=11
\ifx\undefined\newsymbol \else \begingroup\def\input#1 {\endgroup}\fi
\expandafter\ifx\csname amssym.def\endcsname\relax \else\endinput\fi
\expandafter\edef\csname amssym.def\endcsname{%
       \catcode`\noexpand\@=\the\catcode`\@\space}
\catcode`\@=11
\def\undefine#1{\let#1\undefined}
\def\newsymbol#1#2#3#4#5{\let\next@\relax
 \ifnum#2=\@ne\let\next@\msafam@\else
 \ifnum#2=\tw@\let\next@\msbfam@\fi\fi
 \mathchardef#1="#3\next@#4#5}
\def\mathhexbox@#1#2#3{\relax
 \ifmmode\mathpalette{}{\m@th\mathchar"#1#2#3}%
 \else\leavevmode\hbox{$\m@th\mathchar"#1#2#3$}\fi}
\def\hexnumber@#1{\ifcase#1 0\or 1\or 2\or 3\or 4\or 5\or 6\or 7\or 8\or
 9\or A\or B\or C\or D\or E\or F\fi}
\font\tenmsa=msam10
\font\sevenmsa=msam7
\font\fivemsa=msam5
\newfam\msafam
\textfont\msafam=\tenmsa
\scriptfont\msafam=\sevenmsa
\scriptscriptfont\msafam=\fivemsa
\edef\msafam@{\hexnumber@\msafam}
\mathchardef\dabar@"0\msafam@39
\def\maltese{{\mathhexbox@\msafam@7A}}
\font\tenmsb=msbm10
\font\sevenmsb=msbm7
\font\fivemsb=msbm5
\newfam\msbfam
\textfont\msbfam=\tenmsb
\scriptfont\msbfam=\sevenmsb
\scriptscriptfont\msbfam=\fivemsb
\edef\msbfam@{\hexnumber@\msbfam}
\def\Bbb#1{{\fam\msbfam\relax#1}}
\def\widehat#1{\setbox\z@\hbox{$\m@th#1$}%
 \ifdim\wd\z@>\tw@ em\mathaccent"0\msbfam@5B{#1}%
 \else\mathaccent"0362{#1}\fi}
\def\widetilde#1{\setbox\z@\hbox{$\m@th#1$}%
 \ifdim\wd\z@>\tw@ em\mathaccent"0\msbfam@5D{#1}%
 \else\mathaccent"0365{#1}\fi}
\font\teneufm=eufm10
\font\seveneufm=eufm7
\font\fiveeufm=eufm5
\newfam\eufmfam
\textfont\eufmfam=\teneufm
\scriptfont\eufmfam=\seveneufm
\scriptscriptfont\eufmfam=\fiveeufm
\def\frak#1{{\fam\eufmfam\relax#1}}

\csname amssym.def\endcsname
\relax
\newsymbol\smallsetminus 2272
\noblackbox
\newcount\figno
\figno=0
\def\mathrm#1{{\rm #1}}
\def\fig#1#2#3{
\par\begingroup\parindent=0pt\leftskip=1cm\rightskip=1cm\parindent=0pt
\baselineskip=11pt
\global\advance\figno by 1
\midinsert
\epsfxsize=#3
\centerline{\epsfbox{#2}}
\vskip 12pt
\centerline{{\bf Figure \the\figno} #1}\par
\endinsert\endgroup\par}
\font\tenmsb=msbm10       \font\sevenmsb=msbm7
\font\fivemsb=msbm5       \newfam\msbfam
\textfont\msbfam=\tenmsb  \scriptfont\msbfam=\sevenmsb
\scriptscriptfont\msbfam=\fivemsb
\def\Bbb#1{{\fam\msbfam\relax#1}}

\def\Rop{{\Bbb R}}
\def\Zop{{\Bbb Z}}

\def\bbbc{{\mathchoice {\setbox0=\hbox{$\displaystyle\rm C$}\hbox{\hbox
to0pt{\kern0.4\wd0\vrule height0.9\ht0\hss}\box0}}
{\setbox0=\hbox{$\textstyle\rm C$}\hbox{\hbox
to0pt{\kern0.4\wd0\vrule height0.9\ht0\hss}\box0}}
{\setbox0=\hbox{$\scriptstyle\rm C$}\hbox{\hbox
to0pt{\kern0.4\wd0\vrule height0.9\ht0\hss}\box0}}
{\setbox0=\hbox{$\scriptscriptstyle\rm C$}\hbox{\hbox
to0pt{\kern0.4\wd0\vrule height0.9\ht0\hss}\box0}}}}
\def\figlabel#1{\xdef#1{\the\figno}}

\def\pmb#1{\setbox0=\hbox{#1}%
 \kern-.025em\copy0\kern-\wd0
 \kern.05em\copy0\kern-\wd0
 \kern-.025em\raise.0433em\box0 }

\def\half{{1\over 2}}



\def\hsmallsetminus{\hbox{\raise1.5pt\hbox{$\smallsetminus$}}}
\def\tilM{\hbox{${\scriptstyle \widetilde{\phantom M}}$}\hskip-9pt
   \raise1.3pt\hbox{${\scriptstyle M}$}\,}
\def\ie{{\it i.e.}}

%


\def\figin{\epsfcheck\figin}\def\figins{\epsfcheck\figins}
\def\epsfcheck{\ifx\epsfbox\UnDeFiNeD
\message{(NO epsf.tex, FIGURES WILL BE IGNORED)}
\gdef\figin##1{\vskip2in}\gdef\figins##1{\hskip.5in}
\else\message{(FIGURES WILL BE INCLUDED)}%
\gdef\figin##1{##1}\gdef\figins##1{##1}\fi}
\def\DefWarn#1{}
\def\figinsert{\goodbreak\midinsert}
\def\ifig#1#2#3{\DefWarn#1\xdef#1{fig.~\the\figno}
\writedef{#1\leftbracket fig.\noexpand~\the\figno}%
\figinsert\figin{\centerline{#3}}\medskip\centerline{\vbox{\baselineskip12pt
\advance\hsize by -1truein\noindent\footnotefont{\bf Fig.~\the\figno:} #2}}
\bigskip\endinsert\global\advance\figno by1}
\figno=1


\def\II{{\rm II}}

\def\ie{{\it i.e.}}

\def\g{{\frak g}}
\def\p{{\frak p}}


\lref\pwone{P. West, {\it Hidden superconformal symmetry in M-theory},
J. High Energy Phys. {\bf 0008}, 007 (2000); {\tt hep-th/0005270}.}

\lref\pwtwo{P. West, {\it $E_{11}$ and M-theory}, Class. Quant. Grav. 
{\bf 18}, 4443 (2001); {\tt hep-th/0104081}.}

\lref\go{P. Goddard, D.I. Olive, {\it Algebras, 
lattices and strings}, in: {\it Vertex operators in Mathematics and
Physics}, MSRI  Publication $\sharp 3$, Springer (1984) 51.}

\lref\snw{I. Schnakenburg, P. West, {\it Kac-Moody symmetries of IIB
supergravity}, Phys. Lett. {\bf B517}, 421 (2001); 
{\tt hep-th/0107181}.}

\lref\witten{E. Witten, {\it String theory dynamics in various
dimensions}, Nucl. Phys. {\bf B443}, 85 (1995).}

\lref\schwarz{J.H. Schwarz, {\it An SL(2,Z) multiplet of Type IIB
superstrings},  Phys. Lett. {\bf B360}, 13 (1995), Erratum-ibid. 
{\bf B364}, 252 (1995); {\tt hep-th/9508143}.}

\lref\cs{J.H. Conway, N.J.A. Sloane, {\it Sphere packings,
 lattices and groups}, 3rd ed.\ Springer (1999).}

\lref\gow{M.R. Gaberdiel, D.I. Olive, P.C. West, {\it  A class of
Lorentzian Kac-Moody algebras}, {\tt hep-th/0205068}.}

\lref\DDF{E. Del Guidice, P. Di Vecchia, S. Fubini, {\it General
properties of the dual resonance model}, Ann. Phys. {\bf 70}, 378
(1972).} 

\lref\pcwrev{P. West, {\it A brief review of the group theoretic
approach to string theory}, Nucl. Phys {\bf 5B} (Proc. suppl.), 217 
(1988).} 

\lref\vafa{C. Vafa, {\it Evidence for F theory},  Nucl. Phys. 
{\bf B469}, 403 (1996); {\tt hep-th/9602022}.} 

\lref\inv{A. Iqbal, A. Neitzke, C. Vafa, {\it A mysterious duality}, 
{\tt hep-th/0111068}.}

\lref\julia{P. Henry-Labordere, B. Julia, L. Paulot, {\it Borcherds
symmetries in M theory}, J. High Energy Phys. {\bf  0204}, 049 (2002);
{\tt hep-th/0203070}.} 

\lref\lw{N. Lambert, P. West, {\it Coset symmetries in dimensionally
reduced bosonic string theory}, Nucl. Phys. {\bf B615}, 117 (2001);
{\tt hep-th/0107209}.}

\lref\ens{F. Englert, H. Nicolai, A. Schellekens, {\it Superstrings
from $26$ dimensions}, Nucl. Phys. {\bf B274}, 315 (1986).}

\lref\borcherds{R.E. Borcherds, {\it Generalised Kac-Moody algebras},
J. Alg. {\bf 115}, 501 (1988).}

\lref\eten{O. B\"arwald, R.W. Gebert, H. Nicolai, {\it On the
imaginary simple roots of the Borcherds algebra $g_{\II_{9,1}}$}, 
Nucl. Phys. {\bf B510}, 721 (1998); {\tt hep-th/9705144}.}

\lref\etenp{R.W. Gebert, H. Nicolai, P. West, {\it Multistring
vertices and hyperbolic Kac-Moody algebras}, Int. Journ. Mod. Phys. 
{\bf A11}, 429 (1996); {\tt hep-th/9505106}.}

\lref\kac{V.G. Kac, {\it Simple graded algebras of finite growth},
Funct. Anal. Appl. {\bf 1}, 328 (1967).}

\lref\moody{R.V. Moody, {\it Lie algebras associated with generalized
Cartan matrices}, Bull. Amer. Math. Soc. {\bf 73}, 217 (1967).}

\lref\cw{C. Campbell, P. West, {\it $N=2$ $D=10$ nonchiral
supergravity and its spontaneous compactification}, Nucl. Phys.
{\bf B243}, 112 (1984).}

\lref\hn{M. Huq, M. Namazie, {\it Kaluza-Klein supergravity in ten
dimensions}, Class. Quant. Grav. {\bf 2}, 293 (1985).}

\lref\gp{F. Giani, M. Pernici, {\it $N=2$ supergravity in ten
dimensions}, Phys. Rev. {\bf D30}, 325 (1984).}

\lref\sw{J. Schwarz, P. West, {\it Symmetries and transformation of
chiral $N=2$ $D=10$ supergravity}, Phys. Lett. {\bf 126B}, 301 (1983).}

\lref\hw{P.S. Howe, P.C. West, {\it The complete N=2, d = 10
supergravity}, Nucl. Phys. {\bf B238}, 181 (1984).}

\lref\schwarzo{J.H. Schwarz, {\it Covariant field equations of chiral
N=2 D= 10 supergravity}, Nucl. Phys. {\bf B226}, 269 (1983).}

\lref\cjs{E. Cremmer, B. Julia, J. Scherk, {\it Supergravity theory in
eleven dimensions}, Phys. Lett. {\bf 76B}, 409 (1978).}

\lref\pwthree{P. West, {\it Physical states and string symmetries}, 
Mod. Phys. Lett. {\bf A10}, 761 (1995); {\tt hep-th/9411029}.}

\lref\m{G. Moore, {\it Symmetries of the bosonic string S-matrix}, 
{\tt hep-th/9310026} and {\tt hep-th/9404025}.}


\Title{\vbox{\baselineskip12pt
\hbox{hep-th/0207032}
\hbox{KCL-MTH-02-16}}}
{\vbox{\centerline{Kac-Moody algebras in perturbative string theory}}}
\smallskip
\centerline{Matthias R.\ Gaberdiel\footnote{$^\star$}{{\tt
e-mail: mrg@mth.kcl.ac.uk}} and 
Peter C.\ West\footnote{$^{\ddagger}$}{{\tt
e-mail: pwest@mth.kcl.ac.uk}}}
\bigskip
\centerline{\it Department of Mathematics, King's College London}
\centerline{\it Strand, London WC2R 2LS, U.K.}
\smallskip
\vskip2cm
\centerline{\bf Abstract}
\bigskip
\noindent
The conjecture that M-theory has the rank eleven Kac-Moody
symmetry $e_{11}$ implies that Type IIA and Type IIB string
theories in ten dimensions possess certain infinite dimensional
perturbative symmetry algebras that we determine. This prediction is
compared with the symmetry algebras that can be constructed in
perturbative string theory, using the closed string analogues of the
DDF operators. Within the limitations of this construction close
agreement is found. We also perform the analogous analysis for the
case of the closed bosonic string. 

\Date{July 2002}

\newsec{Introduction}

During the last century our understanding of particle physics has been 
transformed by the introduction of group theory. Indeed we are
now sure that all the forces of nature, except for gravity, are
described by Yang-Mills gauge theories based on finite dimensional Lie
groups. At low energies these symmetries are spontaneously broken or
confined, and are therefore not obviously visible. A significant step
towards the identification of the underlying symmetries was the
demonstration that if a global symmetry $G$ is spontaneously broken,
then the resulting massless modes are very often described 
by a non-linear realisation based on $G$, with a subgroup $H$ that is 
preserved by the symmetry breaking mechanism. Thus although one might
not know the underlying theory and the mechanism of symmetry breaking,
one could find the groups involved by comparing the known data with
the results of the theory of non-linear realisations.

Gravity is described, at low energies, by Einstein's theory which,
unlike the other forces, is not a Yang-Mills gauge theory. We now
believe that a consistent theory of gravity requires supersymmetry and
string theory as essential ingredients. Our belief in this is supported
by the realisation that, if one adopts this view point, gravity
becomes automatically unified with the other forces. The relevant
maximal supergravity theories, \ie\ the IIA \refs{\cw,\hn,\gp} and 
IIB \refs{\sw,\hw,\schwarzo} supergravity theories in ten dimensions, 
are uniquely determined by supersymmetry. They define the complete low
energy effective theory of type IIA and IIB superstring theory.  In
turn, the two string theories are believed to be suitable limits of
M-theory which has eleven-dimensional supergravity as its low energy
effective  action \refs{\cjs}. Given the  important role of symmetry
in particle physics it would be reasonable to suppose that M-theory
should also possess a very large symmetry group.   

The symmetry algebras used in particle physics are based on finite
dimensional Lie algebras. More recently, the discovery of Kac-Moody 
and Borcherds algebras \refs{\kac,\moody,\borcherds} has considerably 
enlarged the class of Lie algebras beyond that previously
considered. Although the vertex operator technology of string theory
has played a significant part in the mathematical development of these 
algebras, only a  subset of these, affine Kac-Moody algebras, have played
an important role in string theory and conformal field theory so far. 
\smallskip

Recently it has been shown \refs{\pwone,\pwtwo} that the bosonic
sectors, including gravity, of all maximal supergravity theories can
be described by a non-linear realisation. Unlike the usual treatments
of gauge fields and gravity, this procedure treats the different
fields on the same footing, namely as Goldstone bosons for a certain
symmetry group. The nature of this construction points to the 
possibility that there exist alternative formulations of these
supergravity theories that are invariant under a Kac-Moody (and
possibly a Borcherds) algebra. Although this was not proved in
reference \refs{\pwtwo}, it was possible to show that, if this were
the case, then the symmetry must include the Kac-Moody algebra
$e_{11}$ for the case of eleven dimensional supergravity and IIA
supergravity.  Furthermore, it was shown that the corresponding
Kac-Moody algebra for the IIB theory was also $e_{11}$
\refs{\snw}. This is consistent with the idea that these different
string theories are part of a single theory, namely M-theory, that
also has $e_{11}$ as a symmetry. The different string theories
correspond then to different choices of a `vacuum' in this theory,
together with different subgroups that preserve the vacuum under
consideration.  

In all cases, the $e_{11}$ symmetry contains the  Lorentz group which
is evidently a symmetry of perturbation theory. However, it  also
contains symmetries which are non-perturbative. The simplest example
is provided by the case of IIB  where the $e_{11}$ symmetry contains
the sl(2) symmetry \refs{\sw} of IIB supergravity. Since this symmetry
acts on the dilaton, it mixes perturbative with non-perturbative
phenomena. The $e_{11}$ symmetries therefore also contain in general
a mixture of perturbative and non-perturbative symmetries.  

In the examples of non-linear realisations found in particle physics,
only the preferred subgroup is linearly realised in the non-linear 
realisation. However, in these examples the non-linear realisation
only arises as a low energy effective action of some more fundamental
theory, and in this fundamental theory the full symmetry group is
linearly realised. This gives rise to the expectation that in M-theory,
the $e_{11}$ algebra is actually linearly realised. If this is the
case, then a suitable subalgebra of this symmetry will also be
realised  linearly in string theory.
\medskip

In this paper we want to explore some of the consequences of the
conjecture that M-theory is invariant under the $e_{11}$ symmetry. In
particular, we want to determine the perturbative symmetries of the
IIA and IIB superstring theories that are predicted by this
assumption. Since the symmetry breaking of the $e_{11}$ algebra is
different for the two  theories, we find different perturbative
subalgebras in the two cases. We shall also perform the analogous
analysis for the closed bosonic string that has been proposed to
possess the symmetry $k_{27}$ \refs{\pwtwo}. 

If these symmetries are indeed linearly realised in M-theory (or the
full closed bosonic string theory), the perturbative subalgebras
should also be linearly realised in the appropriate string theories,
and should therefore be accessible. In order to analyse whether these
symmetries are indeed present, we perform a DDF construction for the
different closed string theories. Unlike the open string case where
single DDF operators act on the space of physical states, individual
DDF operators do not preserve the space of physical states of the
closed string. However, one can construct bilinear combinations of DDF 
operators that map physical states to physical states. These bilinear
combinations define symmetries of the string scattering amplitudes,
and they generate certain affine Kac-Moody algebras. The idea that DDF
operators may be important for the construction of symmetries in
string theory was first suggested in \refs{\pwthree}.

The algebras that are generated by these bilinear combinations of DDF
operators can be determined for the different cases, and we find 
considerable agreement with the above predictions. We
regard this as good evidence for the conjecture that M-theory is
$e_{11}$ invariant and that the closed bosonic string has a $k_{27}$
symmetry. 
\bigskip

The paper is organised as follows. In section~2 we determine the
perturbative subalgebras of the full symmetry algebras for the
different cases. The DDF construction for the closed bosonic string,
and the determination of the corresponding symmetry algebra is
performed in section~3. Section~4 deals with the generalisation of
this construction to the supersymmetric case. Finally, in section~5
we explain that the full symmetry algebras that arise are of a rather
special kind that is associated with even self-dual lattices, and we 
speculate about the possibility of an $18$-dimensional string
theory. There is one short appendix that describes our conventions for
the simple roots of $e_8$.

\newsec{Perturbative symmetries} 

It was argued in \refs{\pwtwo} that M-theory possesses a large
symmetry that contains $e_{11}$. As is well known, the circle
compactification of M-theory is described by Type IIA string theory
\refs{\witten}, while Type IIB can be obtained upon compactifying
M-theory on a torus and taking the volume of the torus to zero size
\refs{\schwarz}. From the point of view of Type IIA 
and Type IIB string theory, the symmetries described by $e_{11}$
combine perturbative as well as non-perturbative symmetries. In this
section we want to determine the {\it perturbative} symmetries of Type
IIA and Type IIB that are predicted in this way. As we shall see, a
large part of these symmetries can be understood from a perturbative
string construction that we shall present in later sections. We
shall also carry out the analogous calculation for the closed bosonic
string which was argued to possess a $k_{27}$ symmetry. 

In closed string theory the string coupling constant $g_s$ is related
to the expectation value of the dilaton field, $\phi$, as 
$g_s= \exp \phi$. The perturbative symmetries are therefore 
characterised by the property that they do not transform the dilaton
field. In the approach of reference \refs{\pwtwo} the supergravity
theories are constructed as a non-linear realisation. Every field is
therefore a Goldstone field and so is associated with a generator in
$\g=e_{11}$ or $\g=k_{27}$. Let us denote the generator associated
with the dilaton by $R_I$ where $I=A$ for the case of Type IIA, $I=B$
for Type IIB and $I=CB$ for the closed bosonic string. Within this 
framework, the perturbative symmetry algebra ${\frak p}_I$ is then the 
subalgebra of $\g$ that commutes with the generator $R_I$, modulo
transformations proportional to $R_I$; more specifically 
$$
{\frak p}_I =\{a \in \g \, :  [a,R_I]=0\} / R_I \,.
\eqno(2.1)$$
The bracket on $\p_I$ is that inherited from $\g$. The explicit
expression for the  generator $R_I$ is different for the three cases,  
and we shall therefore consider them in turn.

\subsec{The case of Type IIA}

As was shown in \refs{\pwtwo} Eq.\ (4.21) (see also \refs{\pwone}),
the generator associated with the Type IIA dilaton is given by    
$$ R_A = {1\over 12} 
\left( - \sum_{a=1}^{10} K^a{}_a + 8 K^{11}{}_{11}\right)\,.  
\eqno(2.2)$$
This generator is an element of the Cartan subalgebra of $e_{11}$ and
can be written in terms of the Cartan generators as 
$$
R_A = {1\over 4} H_{8} - {1\over 4} H_6 - {1\over 2} H_{7}\,, 
\eqno(2.3)$$
where we have adopted the numbering of \refs{\gow} that is illustrated
in figure~1.
\ifig\dynkinpic{The Dynkin diagram of 
$e_{11}$. We have adopted the conventions of \refs{\gow}, describing
$e_{11}$ as the very-extended $e_8$ 
algebra.}{\epsfxsize3.5in\hskip-.5cm\epsfbox{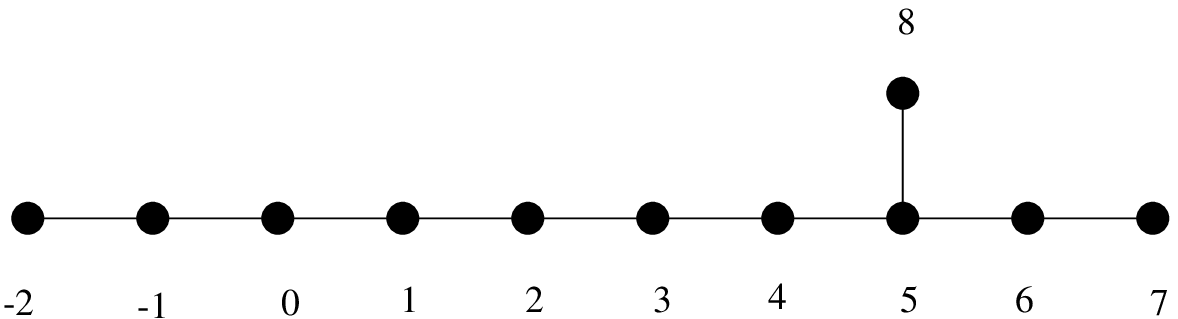}} 

\noindent Given (2.3), the commutators of $R_A$ with the positive
simple roots of $e_{11}$ are 
$$ \eqalign{[R_A, E_a] & = 0\,, \qquad\qquad a=-2,\ldots, 6\,, \cr
[R_A, E_{7}] & = -{3\over 4}\, E_{7}\,, \cr
[R_A, E_{8}] & = {1\over 2}\, E_{8}\,,}
\eqno(2.4)$$
together with the same relations (with opposite signs) for the
negative simple roots $F_a$. The remaining roots of the Kac-Moody
algebra $e_{11}$ are given by  
the multiple commutators 
$$[E_{a_1},[\ldots [E_{a_{p-1}},E_{a_{p}}]\ldots ]\,, 
\eqno(2.5)$$
subject to the Serre relations, as well as similar multiple
commutators of the $F_a$. The multiple commutator in (2.5) describes
the positive root
$$
\sum_{a=-2}^8 n_a\, \alpha_a\,,
\eqno(2.6)
$$
where $\alpha_a$ denotes the root corresponding to $E_a$, and $n_a$
is the non-negative integer that describes the multiplicity with which 
$E_a$ appears in (2.5). The root corresponding to the multiple 
commutator of the $F_a$ is also described by (2.6), except that now
all $n_a$ are non-positive integers. 

The roots of the perturbative subalgebra $\p_A$ are the subset of the
roots in (2.6) for which $(-3/4)\ n_7 + (1/2)\ n_8 = 0$, as follows
from (2.4). The positive roots of $\p_A$ therefore consist of the
roots of $e_{11}$ that are of the form
$$
\sum _{a=-2}^{6} n_a\, \alpha_a 
+m\, (2\, \alpha_{7}+3\, \alpha_{8})\,,
\eqno(2.7)
$$
where $n_a,m\in \Zop_{\geq 0}$, while the negative roots are given by
(2.7) with $n_a,m\in\Zop_{\leq 0}$. 

In order to discover what this algebra actually is, it is useful to
express the roots (2.7) in terms of a set of simple roots. If we
define 
$$
\beta_A=\alpha_{1}+2\alpha_{2}+3\alpha_{3}+4\alpha_{4}+5\alpha_{5}
+3\alpha_{6}+2\alpha_{7}+3 \alpha_{8}\,,
\eqno(2.8)
$$
then $\beta_A^2=2$, $\beta_A.\alpha_{0}=-1$, $\beta_A.\alpha_{6}=-1$,
while $\beta_A . \alpha_a$ with $a=-2,-1,1,2,\ldots,5$ vanish. Thus we
can choose the set of simple roots to consist of 
$$\alpha_{a}\,, \quad a=-2,\ldots, 6\,, \qquad \qquad 
\hbox{and} \qquad\qquad  \beta_A\,.
\eqno(2.9)$$
The corresponding Dynkin diagram is given in figure~2; it is precisely
the Dynkin diagram of the very-extended $su(8)$ Kac-Moody algebra
\refs{\gow}.\footnote{$^\star$}{To every finite dimensional simple Lie
algebra $\frak g$ of rank $r$, one can construct a Lorentzian
Kac-Moody algebra of rank $r+3$. The resulting Lie algebra is called
the very-extended algebra of $\frak g$; further details can be found
in \refs{\gow}.}  The perturbative subalgebra contains the sub-algebra
generated by the simple roots $\alpha_a$, $a=-2,\ldots, 6$ which
corresponds to the SL(10) subgroup in the non-linear realisation whose
Goldstone boson is the graviton.    
\ifig\dynkinpic{The Dynkin diagram of the perturbative $\p_A$
algebra that is isomorphic to the very-extended $su(8)$ Kac-Moody   
algebra.}{\epsfxsize3.0in\hskip-.5cm\epsfbox{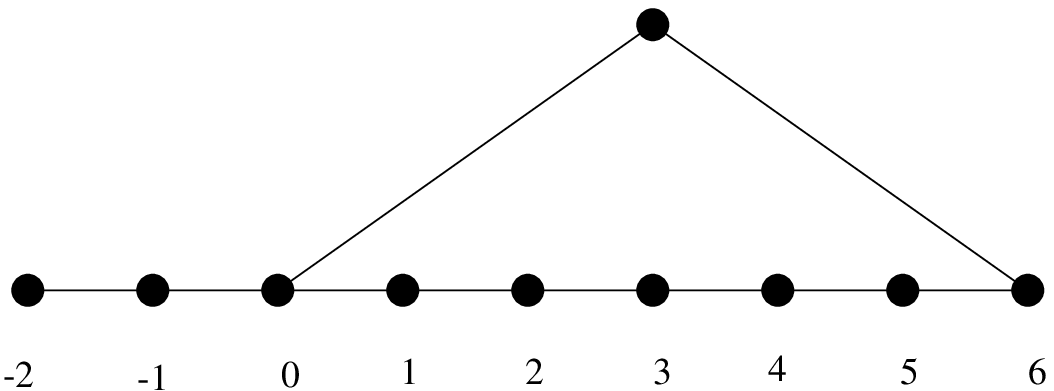}} 
\smallskip

We now give an alternative derivation of $\p_A$, using the description
of the root lattice of $e_{11}$ given in \refs{\gow}. For this it is
useful to consider the Cartan-Weyl basis of the algebra. Recall that
the Cartan generators in the Chevalley and the Cartan-Weyl basis are
related as  $H_a={2\over (\alpha_a,\alpha_a) }\alpha_a^i H_i$, and
that in the Cartan-Weyl basis the generator associated with the root
$\alpha$ obeys  $[H_i,E_\alpha]=\alpha_i E_\alpha$. In terms of the
Cartan-Weyl basis of the algebra, the generator associated with the
dilaton field, given in equation (2.3), can therefore be expressed as  
$R_A= (\alpha_A)^i H_i$, where 
$$
\alpha_A = {1\over 4} \alpha_8 - {1\over 4} \alpha_6 
- {1\over 2} \alpha_7\,. 
\eqno(2.10)$$
The roots of $\p_A$ are then the roots of $e_{11}$ that are orthogonal
to $\alpha_A$.  

Next we use the fact that the root lattice, $\Lambda_{e_{11}}$, of
$e_{11}$ is the lattice
$$
\Lambda_{e_{11}} = \Lambda_{e_8} \oplus \II^{1,1} \oplus
\left\{ x\in \II^{1,1} : x.s = 0 \right\}\,,
\eqno(2.11)
$$
where $s$ is a time-like vector in the second $\II^{1,1}$. Since
$\alpha_A\in\Lambda_{e_8}$, the root lattice of $\p_A$ is then 
$$
\Lambda_{\p_A} = \left\{ x\in\Lambda_{e_8} : x.\alpha_A = 0 \right\} 
\oplus \II^{1,1} \oplus \left\{ x\in \II^{1,1} : x.s = 0 \right\}\,.
\eqno(2.12)
$$
The first lattice of the right hand side is actually the root lattice
of the subalgebra of $e_8$ that commutes with $R_A\in e_8$; in
particular, it is therefore clear that $\p_A$ is a very-extended
Kac-Moody algebra \refs{\gow}. In order to determine what the
commutant of $R_A$ in $e_8$ is, we observe that in the basis of
appendix~A,  
$$
2\, \alpha_A = 
\left({1\over 2},-{1\over 2},-{1\over 2},-{1\over 2},
-{1\over 2},-{1\over 2},-{1\over 2},-{1\over 2}\right)\,.
\eqno(2.13)$$
It is now easy to see that the roots in $\Lambda_{e_8}$ that are
orthogonal to $\alpha_A$ are of the form
$$\pm \left(1,1,0^6\right) \,, \qquad 
\pm \left(0,1,-1,0^5\right)\,,
\eqno(2.14)$$
where in the first case the second $1$ can take any of the remaining
seven places, while in the second case both $1$ and $-1$ can take any
two of the seven remaining places. It is easy to check that these are
precisely the roots of $su(8)$. Thus we have shown that $\p_A$ is the
very-extended $su(8)$ algebra, in agreement with the derivation
above.

\subsec{The case of Type IIB}

The analysis for the case of Type IIB is completely analogous, and we
shall therefore be rather brief. The generator corresponding to the
Type IIB dilaton was determined in \refs{\snw}, and it is given in
terms of the roots of $e_{11}$ as $R_B=H_7$ (in the Chevalley basis)
or as $\alpha_B=\alpha_7$ (in the Cartan-Weyl basis). In terms of the
basis for $e_8$ of appendix~A, we therefore have 
$$\alpha_B = \alpha_7 = \left({1\over 2},{1\over 2},{1\over 2},
{1\over 2}, {1\over 2},{1\over 2},{1\over 2},{1\over 2}\right)\,. 
\eqno(2.15)$$
Using a similar analysis as in the case of IIA one finds that the
subalgebra of $e_8$ that commutes with $\alpha_B$ is actually
$e_7$.\footnote{$^\dagger$}{This also follows from the statement on
page 124 of \refs{\cs}.} Since $\alpha_B\in e_8$, it then follows, as 
above, that $\p_B$ is the very-extended Kac-Moody algebra
corresponding to $e_7$. More specifically, we can choose the simple
roots of $\p_B$ to be given by 
$$\alpha_{a}\,, \quad a=-2,\ldots, 5,8\,, \qquad \qquad 
\hbox{and} \qquad\qquad  \beta_B\,,
\eqno(2.16)$$
where 
$$
\beta_B = \alpha_4 + 2 \alpha_5 + 2 \alpha_6 + \alpha_7 + \alpha_8 \,.
\eqno(2.17)
$$
It is easy to see that $\beta_B^2=2$, $\beta_B.\alpha_3=-1$, while 
$\beta_B.\alpha_a=0$ for $a=-2,-1,\ldots,2,4,5,8$. The corresponding
Dynkin diagram is shown in figure~3.
\ifig\dynkinpic{The Dynkin diagram of the perturbative $\p_B$
algebra that is isomorphic to the very-extended $e_7$ Kac-Moody   
algebra.}{\epsfxsize3.0in\hskip-.5cm\epsfbox{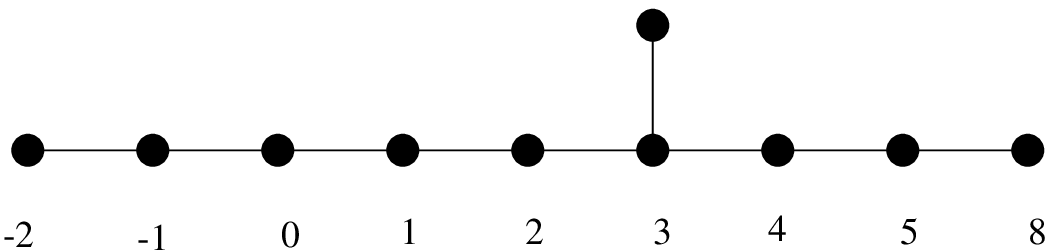}}

\subsec{Some comments}

It is also instructive to determine the symmetry algebra that is
simultaneously perturbative with respect to both Type IIA and Type
IIB. The subalgebra of $e_8$ that commutes with both $\alpha_A$ and
$\alpha_B$ is actually equal to $su(7)$, and thus it follows that the
common perturbative subalgebra is the very-extended algebra of
$su(7)$. This Lie algebra contains $su(9)$, which has a natural
interpretation in terms of the compactification of M-theory on
$T^2\times T^9$. If we are interested in symmetries that preserve the
dilaton of IIA and IIB, this implies that the moduli of the $T^2$ are
frozen. The remaining perturbative symmetries are then those of a
$9$-torus, thus giving rise to SL$(9,\Zop)$; the corresponding
continuous group is thus SU(9).  

It is amusing to observe that $e_{11}$ sits rather naturally inside
$\II^{10,2}$ (see (2.11) above). The lattice  $\II^{10,2}$ is
obviously reminiscent of F-theory \refs{\vafa}. From this point of
view the direction corresponding to the vector $s$ in (2.11) plays the
role of the time-like direction on which one has to compactify
F-theory in order to obtain M-theory. It is also somewhat suggestive
that the vectors (2.13) and (2.15), whose orthogonal complement
define the perturbative subalgebras for IIA and IIB, respectively,
are `spinor weights' of opposite chirality. 
It would be very interesting to understand these observations more
conceptually. It would also be interesting to understand the relation
of the $e_{11}$ symmetry of M-theory to those obtained recently in 
\refs{\inv,\julia}.

\subsec{The closed bosonic string}

The low energy effective action for the closed bosonic string is
thought to be invariant under the Kac-Moody algebra $k_{27}$ of rank
$27$ \refs{\pwtwo}. Its Dynkin diagram is shown in figure~4. 
\ifig\dynkinpic{The Dynkin diagram of the algebra $k_{27}$ 
that is isomorphic to the very-extended $d_{24}$ Kac-Moody   
algebra.}{\epsfxsize4.8in\hskip-.5cm\epsfbox{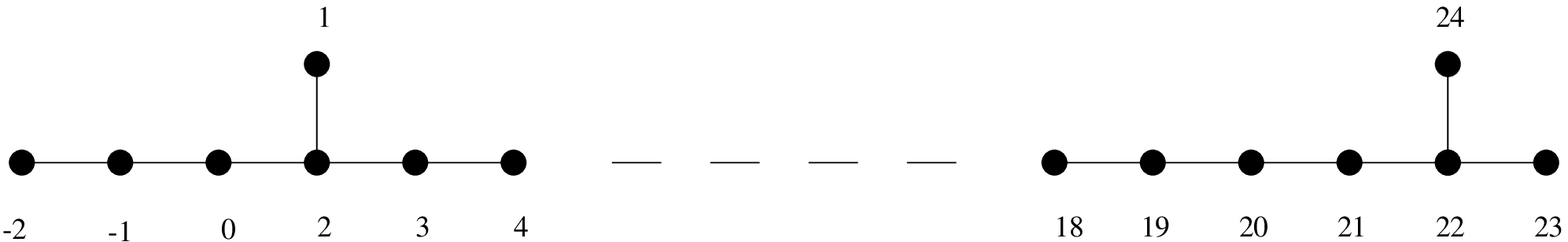}} 
The dilaton is the Goldstone boson which is associated to the
generator $R_{CB}$ that is an element of the Cartan subalgebra. Its
commutation relations with the Chevalley generators of $k_{27}$ are
given as \refs{\pwtwo} 
$$ \eqalign{
[R_{CB}, E_a] & = 0\,, \qquad\qquad a=-2,-1,0,2,3,\ldots, 23\,,\cr
[R_{CB}, E_{1}] & = - E_{1}\,, \cr 
[R_{CB}, E_{24}] & = E_{24}\,,}
\eqno(2.18)
$$
together with the same relations (with the opposite signs) with $E_a$
being replaced by $F_a$. Using the same arguments as above for the
case of Type IIA, the positive roots of $\p_{CB}$ then consist of
multiple commutators of the $E_a$ that contain $E_{1}$ and $E_{24}$
with the same multiplicity. If we define 
$$
\beta_{CB}=\sum_{a=2}^{22} \alpha_{a} + \alpha_1 + \alpha_{24}\,,
\eqno(2.19)$$
then $\beta^2=2$, $\beta.\alpha_{0}=-1$, $\beta.\alpha_{23}=-1$ and 
$\beta.\alpha_{a}=0$ for $a=-2,-1,2,3,\ldots, 22$. As such,  
$$\alpha_{a}\,, \quad a=-2,-1,0,2,3,\ldots, 23\,, \qquad \qquad
\hbox{and}\qquad \qquad \beta  
\eqno(2.20)$$
form a set of simple roots for the perturbative algebra. It is easy to
see that the corresponding Dynkin diagram is that of the very-extended
$a_{23}$ algebra. It contains the subalgebra generated by 
the simple roots $\alpha_a, \ a=-2,-1,0,2,3,\ldots, 23$ which
corresponds to the SL(26) sub in the non-linear realisation
whose Goldstone boson is the graviton.

\newsec{Kac-Moody symmetries in the bosonic string}

We now want to show that the space of states of the closed
string theory naturally carries an action of a certain
Kac-Moody algebra that we shall construct. We shall also explain that
this is a symmetry of the perturbative string scattering
amplitudes. The  construction makes use of the DDF-operators
\refs{\DDF}. We shall first consider the closed bosonic string; the
analogous analysis for the superstring will be described in the next
section.

\subsec{The DDF construction of the open bosonic string}

The DDF operators are usually discussed within the context of the open
(bosonic) string; it will therefore be instructive to review this
construction briefly before generalising it to the closed bosonic
string. The open bosonic string sweeps out a surface in space-time
described by the fields $X^\mu(\tau, \sigma)$, where $\tau$ and
$\sigma$ are the time- and space-parameters of the world-sheet. It is
convenient to rewrite the dynamical variable $X^\mu(\tau,\sigma)$ in
terms of the variables $q^\mu\equiv q^\mu(\tau=0)$ and the oscillator
modes $\alpha_n^\mu \equiv \alpha^\mu_n(\tau=0)$ as  
$$X^\mu(\tau,\sigma)=q^\mu+\sqrt{2\alpha'} \alpha_0^\mu \tau
+i\sqrt{\alpha'\over 2} \sum_{n\ne 0} {1\over n}
\left(\alpha_n^\mu e^{-in(\tau+\sigma)}
+\alpha_n^\mu e^{-in(\tau-\sigma)}\right)\,.
\eqno(3.1)
$$
The space-time momentum $p^\mu$ is given by 
$\alpha_0^\mu=\sqrt{2\alpha'}p^\mu$ and one usually chooses
$2\alpha'=1$. The field $Q^\mu(\tau)=X^\mu(\tau,0)$ can then be
written as   
$$Q^\mu(z)=q^\mu-i p^\mu\ln z
+i \sum_{n\ne 0} {1\over n}\,\alpha_n^\mu \, z^{-n}\,,
\eqno(3.2)
$$
where $z=e^{i\tau}$. 

In terms of the usual light-cone coordinates 
$p^\pm={1\over \sqrt 2}(p^{25}\pm p^{0})$ we then choose a particular
Lorentz frame by taking the tachyonic ground state to have the
momentum $p_0^+=p_0^-=1$, $p_0^i=0$. The transverse DDF operators
\refs{\DDF} are then defined by 
$$
A^i_n={1\over 2\pi i} \oint_0\,  
i\partial Q^i(z) \, e^{i n k_0\cdot Q(z)}\, dz\,, 
\eqno(3.3)
$$
where $k^-_0=1$, and $k^+_0=k^i_0=0$. In writing (3.3) we do not have
to worry about a normal ordering prescription since $k_0^2=0$, and
the spatial components of $k^\mu_0$ vanish. The contour integral is
well-defined provided that $A^i_n$ acts on states whose momentum has
integer inner product with $k_0$. In particular, this condition is
satisfied for the ground tachyon ground state with momentum $p_0$, and
any state that is created from it by the action of the DDF operators
(as the DDF operator $A^i_n$ changes the momentum by $n k_0$).

Using standard techniques it is easy to show that the DDF operators
obey the relations of a $u(1)$ current algebra
$$  
[ A^i_m, A^j_n ]  = m \  \delta^{ij} \ \delta_{m,-n} \,.
\eqno(3.4)$$
The DDF operators commute with the Virasoro generators 
$L_n={1\over 2}\sum_m :\alpha_m\cdot \alpha_{n-m}:$ and, as a result,
we can create physical states by acting with polynomials of the DDF
operators on the tachyonic ground state $|p_0\rangle$. All of these
states are annihilated by $L_n$ with $n>0$ and satisfy the mass-shell
condition $L_0\psi = \psi$. In the critical dimension $26$, all
physical states with momentum $p_0+N k_0$ can be created in this way.

\subsec{The DDF construction of the closed bosonic string}

Let us now generalise the DDF construction to the $26$-dimensional
closed bosonic string theory. The fields $X^\mu$ that describe the
embedding of the string in the space-time can now be written as  
$$X^\mu(\tau,\sigma)=q^\mu+\sqrt{2\alpha'} \alpha_0^\mu\tau
+i\sqrt{\alpha'\over 2} \sum_{n\ne 0} {1\over n}
\left(\alpha_n^\mu e^{-in(\tau+\sigma)}+\bar \alpha_n^\mu
e^{-in(\tau-\sigma)}\right)\,,
\eqno(3.5)$$
where the space-time momentum $p^\mu$ is given by 
$\alpha_0^\mu=\sqrt {\alpha'\over 2}p^\mu$. It is traditional to take
$\alpha'=2$. The physical states $\Psi$ of the theory satisfy the
mass-shell condition,  
$$
L_0 \Psi = \bar{L}_0 \Psi = \Psi\,,
\eqno(3.6)$$
as well as the Virasoro constraints
$$
L_n \Psi = \bar{L}_n \Psi = 0 \qquad \hbox{for all $n>0$.}
\eqno(3.7)$$
Here $L_n$ and $\bar L_n$ are defined as 
$L_n={1\over 2}\sum_m :\alpha_m\cdot \alpha_{n-m}:$ 
and $\bar L_n={1\over 2}\sum_m :\bar\alpha_m\cdot \bar\alpha_{n-m}:$. 
In defining $\bar L_n$ we have used the mode $\bar\alpha^\mu_0$,
which at this stage simply equals $\alpha^\mu_0=\bar\alpha^\mu_0$.
The oscillator modes satisfy the commutation relations
$$\eqalign{
{}[\alpha^\mu_m,\alpha^\nu_n] & = m\, \delta^{\mu\nu}\, \delta_{m,-n}\,,
\cr 
{}[\alpha^\mu_m,\bar\alpha^\nu_n] & = 0\,, \cr
{}[\bar\alpha^\mu_m,\bar\alpha^\nu_n] & = m\, \delta^{\mu\nu} \,
\delta_{m,-n}\,.\cr}
\eqno(3.8)
$$

In order to describe the space of states of the closed string it is
convenient to introduce separate left- and right-moving momentum
operators $p_L=\alpha_0$ and $p_R=\bar\alpha_0$ without imposing
initially that they agree, as well as the corresponding position
operators $q_L,q_R$.\footnote{$^\ddagger$}{From a more abstract point of
view this means that we are complexifying the space-time.} These
operators satisfy the commutation relations    
$$
\eqalign{
{} [q_L^\mu,p_L^\nu] & = i \eta^{\mu\nu}\,, \cr 
{} [q_L^\mu,q_R^\nu] & = [q_L^\mu,p_R^\nu] = [q_R^\mu,p_L^\nu] = 
[p_L^\mu,p_R^\nu]= 0\,,
\cr 
{} [q_R^\mu,p_R^\nu] & = i \eta^{\mu\nu}\,,\cr}
\eqno(3.9)
$$
where $\eta^{\mu\nu}$ is the Minkowski metric. As before for the case
of the open string we introduce the fields  
$${\eqalign{
X^\mu_L(z) & = q_L^\mu - i p_L^\mu \ln z + i \sum_{n\ne 0}
{1\over n}\, \alpha^\mu_n \, z^{-n}\,, \cr
X^\mu_R(z) & = q_R^\mu - i p_R^\mu \ln \bar{z} + i \sum_{n\ne 0}
{1\over n}\, \bar\alpha^\mu_n \, \bar{z}^{-n}\,.}}
\eqno(3.10)$$
The field of equation (3.5) can then be expressed as 
$$X^\mu(z,\bar{z})= X^\mu_L(z) + X^\mu_R(\bar{z})\,,
\eqno(3.11)$$
where $z=\tau+\sigma$ and $\bar{z}=\tau-\sigma$ provided we take 
$p_L^\mu=p^\mu_R=p^\mu$ and $q^\mu=q_L^\mu+q_R^\mu$. The actual space
of states is therefore the subspace of our larger space where
we impose the `reality condition' $p_L^\mu=p^\mu_R$. This larger space
is generated by the action of the modes $\alpha^\mu_n$ and
$\bar\alpha^\mu_n$ with $n<0$ on the ground states
$|(p_L,p_R)\rangle$, where $(p_L,p_R)$ denotes the ground state 
momentum, \ie\  
$$p_L^\mu\, | (k_L,k_R) \rangle = k_L^\mu \, |(k_L,k_R)\rangle\,,
\qquad\qquad 
p_R^\mu\, | (k_L,k_R) \rangle = k_R^\mu \, |(k_L,k_R) \rangle
\,.
\eqno(3.12)$$
The ground states are annihilated by the oscillators $\alpha^\mu_n$
and $\bar\alpha^\mu_n$ with $n>0$.

Next we introduce the (chiral) vertex operators
$${\eqalign{
V^i(nk_0,z) & =  i\partial X^i_L(z)\, e^{i n k_0 \cdot X_L(z)}\,,  \cr 
\bar{V}^i(nk_0,\bar{z}) & = i\bar \partial X^i_R(\bar{z}) \,
                 e^{i n k_0 \cdot X_R(\bar{z})}  \,,}}
\eqno(3.13)$$
where $k^-_0=1,k^+_0=0=k^i_0$, and as before no normal ordering
prescription is required. We then define their `zero' mode by 
$${\eqalign{
A^i_n & = {1\over 2\pi i} \oint_0 {dz}\, V^i(nk_0,z)\,, \cr
\bar{A}^i_n & = {1\over 2\pi i} \oint_0 {d\bar{z}} \,
\bar{V}^i(nk_0,\bar{z}) \,.}}
\eqno(3.14)$$
These operators are the closed string analogues of the open
string DDF-operators of equation (3.3). They are well-defined on
states whose left- and right-moving momentum has integer inner product
with $k_0$. 

We observe that the operators $X_L$ and $X_R$ both have the same form
as $Q^\mu$ of equation (3.2), and that the operators they contain
have the same commutation relations as those in $Q^\mu$. As such, they
satisfy simple algebra relations among themselves 
$${\eqalign{
{} [ A^i_m, A^j_n ] & = m \, \delta^{ij}\, \delta_{m,-n}\,, \cr
{} [ A^i_m, \bar{A}^j_n ] & = 0\,, \cr
{} [ \bar{A}^i_m, \bar{A}^j_n ] & = m \, \delta^{ij}\,
\delta_{m,-n}\,, }}
\eqno(3.15)$$
by analogy with equation (3.4). Furthermore, for the same reason, they
commute with the two Virasoro algebras
$${\eqalign{
{} [ L_m , A^i_n] & = 0\,, \qquad \hbox{for all $m,n\in\Zop$}\,, \cr
{} [ \bar{L}_m , A^i_n] & = 0\,, \qquad \hbox{for all $m,n\in\Zop$,}}}
\eqno(3.16)$$
and similarly for $\bar{A}^i_n$. Here we have defined $\bar L_n$ as in
the paragraph following (3.7), but we do not now assume that
$\bar\alpha_0=\alpha_0$. The DDF operators map states satisfying 
equation (3.6) and (3.7) into themselves. However, $A^i_n$ changes the
left-moving momentum $p_L$ to $p_L+n k_0$, while $\bar{A}^i_m$ changes
the right-moving momentum $p_R$ to $p_R+m k_0$. In particular, a
single DDF operator with $n\ne 0$ therefore does not preserve the
condition $p_L=p_R$. This will play an important role in the
following.  

The tachyonic ground state is described by $|(p,p)\rangle$ and it
satisfies all the physical state conditions (and in particular the
mass-shell condition) provided that $p^2=2$. As before we choose a
fixed vector $p_0$ with $p_0^2=2$ by taking $p_0^+=p_0^-=1$ with
$p^i=0$. With this choice $p_0\cdot k_0 = 1$. The DDF operators with
positive moding annihilate the tachyon ground state
$${
A^i_n |(p_0,p_0)\rangle = \bar{A}^i_n |(p_0,p_0)\rangle = 0 \qquad 
\hbox{for $n>0$,}}
\eqno(3.17)$$
since the momentum of these states is $(p_0+nk_0,p_0)$
and $(p_0,p_0+nk_0)$, respectively, and no state with (left- or
right-moving) momentum of the form $p_0+nk_0$ with $n>0$ can satisfy
equation (3.6) as $(p_0+nk_0)^2=2+2n$. 

On the other hand, the DDF operators corresponding to negative
modes do not annihilate the tachyonic ground state, and lead to the
states 
$${
\prod_{k=1}^{r} A^{i_k}_{-n_k} \, \prod_{l=1}^{s} \bar{A}^{j_l}_{-m_l} 
|(p_0,p_0)\rangle \,.}
\eqno(3.18)$$
These states have momentum of the form $(p_0-n_L k_0,p_0-n_R k_0)$, 
where $n_L=\sum_{k=1}^{r} n_k$ and $n_R=\sum_{l=1}^{s} m_l$. In
particular, the action of the closed string DDF operators is therefore
well-defined on these states. Although the resulting states satisfy
the conditions of equations (3.6) and (3.7), they will not in general
be physical states of the closed bosonic string as they do not
necessarily obey $p_L=p_R$. However, they will be physical states  if
they satisfy 
$$n_L=n=n_R\,, \quad \hbox{where}\quad
n_L=\sum_{k=1}^{r} n_k\,, \quad \hbox{and} \quad 
n_R=\sum_{l=1}^{s} m_l\,.
\eqno(3.19)$$
If this is the case, the corresponding state has left- and
right-moving momentum given by $p=p_0 - n\, k_0$ where $p^2=2(1-n)$. 
Furthermore, since we have $\alpha^\mu_0=\bar\alpha^\mu_0$ on the
state, the conditions of equations (3.6) and (3.7) actually imply the
physical state conditions.\footnote{$^\star$}{Initially we only know
that these states satisfy (3.6) and (3.7) where $\bar L_n$ is defined
as in the paragraph following (3.7), but without assuming that 
$\alpha^\mu_0=\bar\alpha^\mu_0$. If we have
$\alpha^\mu_0=\bar\alpha^\mu_0$ on the state, this reproduces then
precisely the actual physical state condition.}

It follows from the commutation relations (3.15) that the
corresponding states have positive norm. It is also easy to see that
in the critical dimension these states account for all the physical
closed string states with momentum $p=p_0 - n\, k_0$. All physical
states (except those with $p=0$) can be related by a Lorentz
transformation to a state of the form of equation (3.18) with
$n_L=n_R$.

\subsec{Symmetries of the spectrum}

In the previous subsection we have given a simple description of the
space of states of the closed bosonic string. Now we want to construct
operators that map this space into itself. As we have mentioned above,
the only single DDF-operators that preserve the level matching
condition of equation (3.19)  are $A^i_0$ and $\bar{A}^i_0$ that act
rather trivially (they are actually equal to $p^i_L$ and
$p^i_R$). Unlike the situation for the open bosonic string, we
therefore have to consider products of DDF-operators in order to
construct non-trivial operators that respect (3.19). The simplest
non-trivial example arises when these operators are bilinear in
$A^i_n$ and $\bar{A}^i_n$. Then there are three different generators
that preserve the level matching condition 
$${
L^{ij}(n) = {1\over n} A^i_{-n} A^j_{n} +\half \delta^{ij}\,,
\qquad 
\bar{L}^{ij} (n) = {1\over n} \bar{A}^i_{-n} \bar{A}^j_{n} 
+ \half \delta^{ij}\,,}
\eqno(3.20)$$
where $n>0$ (the expressions for $n$ and $-n$ differ only by a
constant) as well as 
$${
K^{ij}(m) = {1\over m} A^i_{-m} \bar{A}^j_{-m} \,,}
\eqno(3.21)$$
where $m\ne 0$. The prefactors of $1/n$ and $1/m$, as well as the
terms proportional to $\delta^{ij}$ have been introduced for later
convenience.  

Given the commutation relations of the DDF-operators of equation
(3.15), we can determine the commutation relations of the operators
$L^{ij}(n)$, $\bar{L}^{ij}(n)$ and $K^{ij}(n)$. Let us first consider
the commutator of two $L^{ij}(n)$ operators. It is given as 
$${\eqalign{
{}[L^{ij}(n),L^{kl}(m)] 
& = {1\over nm} [A^i_{-n} A^j_{n},A^k_{-m} A^l_{m}] \cr
& = {1\over nm} \left( A^i_{-n} [A^j_n,A^k_{-m}] A^l_{m} +
            A^k_{-m} [A^i_{-n},A^l_m] A^j_{n} \right) \cr 
& = \delta_{n,m} 
    \left( \delta^{jk} L^{il}(n) - \delta^{il} L^{kj}(n)\right) \,,}}
\eqno(3.22)$$
where we have used that $n,m>0$, and therefore that only two of the
four possible commutator terms can be non-trivial. We recognise these
commutation relations as those of the Lie algebra $u(24)$. Thus, for
fixed $n$, the modes $L^{ij}(n)$ define the Lie algebra of $u(24)$,
while the modes for different $n$ commute. Similarly we find that 
$${
{}[\bar{L}^{ij}(n),\bar{L}^{kl}(m)]  = 
\delta_{n,m} 
\left( \delta^{jk} \bar{L}^{il}(n) 
  - \delta^{il} \bar{L}^{kj}(n)\right)\,.}
\eqno(3.23)$$
and therefore also the modes $\bar{L}^{ij}(n)$ form the
Lie algebra of $u(24)$. Furthermore, the modes $L^{ij}(n)$ and 
$\bar{L}^{ij}(m)$ commute. 

Finally, we find that the commutation relations with the 
$K^{kl}(m)$ generators are given by 
$${\eqalign{
{}[ L^{ij}(n), K^{kl}(m)]  & 
= \delta_{n,m} \delta^{jk} K^{il}(m) 
- \delta_{n,-m} \delta^{ik} K^{jl}(m)\,, \cr
{}[ \bar{L}^{ij}(n),K^{kl}(m)]  & 
= \delta_{n,m} \delta^{jl} K^{ki}(m) 
- \delta_{n,-m} \delta^{il} K^{kj}(m)\,,  \cr
{}[K^{ij}(n),K^{kl}(m)] & =  \delta_{n,-m} 
\left( \delta^{ik} \bar{L}^{jl}(n) + 
       \delta^{jl} L^{ik}(n) \right) \,.}}
\eqno(3.24)$$
In writing  these equations we have assumed, without loss of
generality, that $n>0$ while $m$ was unrestricted.

Incidentally, the operators $L^{ij}(n)$, $\bar{L}^{ij}(n)$ and 
$K^{ij}(m)$ can also be defined in the original space of states, 
without the need of relaxing $p_L=p_R$ (or complexifying
space-time). For example we can define
$$
K^{ij}(m) = {1\over m} \,{1\over 2\pi i} \oint {dz} \, 
{1\over 2\pi i}\oint {d\bar{z}} \,
P^i(z) \, \bar{P}^j(\bar{z}) : \exp(i m k_0 \cdot X(z,\bar{z})) : \,,
\eqno(3.25)
$$
where $P^i(z) = i{\partial \over \partial z} X^i(z,\bar{z})$, 
$\bar{P}^j(z) = i{\partial \over \partial \bar z} X^j(z,\bar{z})$, and  
the coordinates $z$ and $\bar{z}$ are regarded as independent (\ie\
not as complex conjugates of one another). One can show that
$K^{ij}(m)$, as expressed in equation (3.25), commutes with $L_n$ and
$\bar L_n$. Furthermore, this definition of $K^{ij}(m)$ agrees, on the
actual physical subspace, with that given above in (3.21). Similar
formulae also exist for  $L^{ij}(n)$ and $\bar{L}^{ij}(n)$, and the
commutation relations (3.22) -- (3.24) can be derived in this
way. 

It follows from these commutation relations that, for each fixed
$n>0$, the modes 
$$
L^{ij}(n)\,, \qquad \bar{L}^{ij}(n)\,, \qquad 
K^{ij}(n)\,, \qquad K^{ij}(-n)
\eqno(3.26)$$
form a closed subalgebra, and that the algebras corresponding to
different positive values of $n$ commute. For each $n>0$, there are  
$4 (24)^2= (48)^2$ modes in equation (3.26), and in fact the algebra
generated by these modes is precisely $u(48)$. In order to see this we
observe that the adjoint representation of $u(48)$ decomposes with
respect to $u(24)\oplus u(24)$ into the adjoint representation of 
$u(24)\oplus u(24)$, as well as two bi-fundamental
representations. As we have explained above, the modes $L^{ij}(n)$ and 
$\bar{L}^{ij}(n)$ generate the adjoint representation of 
$u(24)\oplus u(24)$, and it follows from (3.24) that both the modes 
$K^{ij}(n)$ and $K^{ij}(-n)$ transform in the
bi-fundamental representation of $u(24)\oplus u(24)$.  

Next we want to show that these infinitely many commuting finite Lie 
algebras actually organise themselves into an affine Kac-Moody
algebra. The construction we now present works for the various
subalgebras, and in particular for $u(48)$ itself, but we give it for
the case of the diagonal $u(24)$ subalgebra of 
$u(24)\oplus u(24)$. The generators $L^{ij}(n)$ and  
$\bar{L}^{ij}(n)$ do not change the momentum of a state on
which they act (and therefore, in particular, do not change their
mass), whereas the generators $K^{ij}(n)$ typically
do. If we are interested in perturbative  symmetries of the theory, we
should therefore only consider the generators $L^{ij}(n)$ and 
$\bar{L}^{ij}(n)$. Furthermore, it is natural to 
consider the left-right symmetric combination of these generators
(that defines the diagonal subalgebra) since this will have a 
simple geometrical interpretation. The modes we consider are therefore 
$$
R^{ij}(n) = L^{ij}(n) + \bar{L}^{ij}(n)\,,
\eqno(3.27)$$
which satisfy the commutation relations 
$$
{}[R^{ij}(n),R^{kl}(m)] = \delta_{n,m} 
\left( \delta^{jk} R^{il}(n) - \delta^{il} R^{kj}(n)\right) ,
\eqno(3.28)$$
as follows from equations (3.24). If we are only interested in these
generators, it is not necessary to introduce the $\delta^{ij}$
correction term in the definition of $L^{ij}$ and $\bar L^{ij}$ in
(3.20) since it drops out of (3.22) and (3.23). 

For each fixed $n$, the modes $R^{ij}(n)$ only `see' the DDF-operators
with mode number $\pm n$; it is therefore natural to consider the
linear combination  
$$
R^{ij}_0 = \sum_{r=1}^{\infty} R^{ij}(r) \,.
\eqno(3.29)$$
This operator then acts completely geometrically on the space-time
indices of the fields and indeed the part which is anti-symmetric in
the $(i,j)$ indices is just the generator of the transverse Lorentz
transformations. We should stress that $R^{ij}_0$ is well-defined on
the (Fock) space of DDF-states since for any given state of finite
momentum, the sum in equation (3.29) terminates. This follows
directly from the commutation relations of the DDF-operators, as well
as the property that the positive DDF-operators annihilate the ground
state.    

We can generalise the generator of equations (3.29)  as follows. Let
$\zeta$ be an arbitrary (complex) number with $\zeta\ne 0,1$. Then we
define for any $m\in\Zop$  
$$
R^{ij}_m = \sum_{r=1}^{\infty} \zeta^{rm} R^{ij}(n)\,.
\eqno(3.30)$$
By the same arguments as above, each of the operators $R^{ij}_m$ is 
well-defined on the Fock space of DDF-states. 

The modes $R^{ij}_n$ satisfy the commutation relations
$${\eqalign{
{}[R^{ij}_m,R^{kl}_n] & = \sum_{r,s=1}^{\infty} \zeta^{rm+sn} 
[R^{ij}(r),R^{kl}(s)] \cr
& = \sum_{r,s=1}^{\infty} \zeta^{rm+sn} \delta_{r,s}
\left[ \delta^{jk} R^{il}(r) - \delta^{il} R^{kl}(r) \right] \cr
& = \sum_{r=1}^{\infty} \zeta^{r(m+n)} 
\left[ \delta^{jk} R^{il}(r) - \delta^{il} R^{kl}(r) \right] \cr
& = \delta^{jk} R^{il}_{m+n} - \delta^{il} R^{kl}_{m+n} \,.}}
\eqno(3.31)$$
These are precisely the commutation relations of the affine Kac-Moody
algebra $\hat{\rm u}$(24) with a vanishing central term. 
The algebra $\hat{\rm u}$(24) is not simple since it can
be decomposed as 
$$
\hat{\rm u}(24) = \hat{\rm su}(24) \oplus \hat{\rm u}(1)\,.
\eqno(3.32)$$
In particular, it therefore contains the simple Kac-Moody algebra
$\hat{su}(24)$. It is worth pointing out that the bilinear
combinations of DDF operators that appear in $R^{ij}_n$ were required
by the level matching condition (3.19), and that the resulting
operators define non-abelian commutation relations; this is in marked
contrast to the open string DDF operators that only satisfy the
relations (3.4). These non-abelian symmetries are thus characteristic
of closed string theories.

The reader may find the occurrence of an arbitrary parameter in the
construction of equation (3.30) rather artificial. However, we note
that taking $\zeta \to \zeta^q$ is equivalent to replacing $R^{ij}_n$
by $R^{ij}_{qn}$. This leads to an affine algebra that is isomorphic
to the original affine algebra. This ambiguity in defining the
generators is always present for an affine algebra without centre, 
and this is responsible for the arbitrariness in choosing $\zeta$.

\subsec{Symmetries of amplitudes}

As we have seen in the previous subsection, the space of states of the
closed bosonic string carries an action of the Kac-Moody algebra
$\hat{u}(24)$. We now want to show that this Kac-Moody algebra acts as
a symmetry on the scattering amplitudes of the closed bosonic string.  
It was argued in reference \refs{\pwthree} that the DDF operators could 
be regarded as symmetry operators in the open bosonic string. Here we 
generalise this consideration  to the closed bosonic string, but unlike   
\refs{\pwthree}, which used the `group theoretic' approach to string
theory reviewed in \refs{\pcwrev}, we will use more conventional 
conformal field theory techniques.

First we recall from the definition (3.14) that each (single)
DDF-operator is the zero-mode of a chiral vertex operator of conformal
weight one. For amplitudes on the Riemann sphere it therefore follows
from usual conformal field theory arguments that 
$$
0 = \sum_{i=1}^n \langle V(\psi_1,z_1)\cdots V(\psi_{i-1},z_{i-1}) 
V(A \psi_i,z_i) V(\psi_{i+1},z_{i+1})\cdots V(\psi_n,z_n)\rangle
\,,
\eqno(3.33)
$$
where $A$ is such a zero mode. In writing (3.33) we have assumed that
the action of $A$ is well-defined on each of the $n$ external
states. This is simply the statement that their momentum along the
light-cone directions lies in the lattice spanned by $p_0$ and
$k_0$. Thus, in effect, we are assuming now that the string theory has
been compactified on the lattice $\II^{1,1}$ in these two directions. 

The generators of the $\hat{u}(24)$ symmetry are bilinears in the DDF
operators. Applying the above argument to each of the two
DDF-operators separately we therefore find that 
$$
\eqalign{
0 = & \sum_{i\ne j} \langle V(\psi_1,z_1)\cdots V(A\psi_{i},z_{i}) 
\cdots V(B \psi_j,z_j) \cdots V(\psi_n,z_n)\rangle \cr 
& \quad + 
\sum_{i=1}^n \langle V(\psi_1,z_1)\cdots V(AB\psi_{i},z_{i}) 
\cdots V(\psi_n,z_n)\rangle\,,}
\eqno(3.34)
$$
where we have written the two single DDF-operators schematically as
$A$ and $B$. Thus if we define the action of $AB$ on the $n$-fold
tensor product of external states as in (3.34), then the amplitudes
are  invariant under this action. In this sense the Kac-Moody algebra
$\hat{u}(24)$ defines a symmetry of the scattering amplitudes. It may
also be worthwhile pointing out that the first sum of (3.34) describes
amplitudes where two of the external states are not on-shell (since
the single DDF operators $A$ and $B$ do not map physical states to
physical states). It is therefore conceivable that these contributions
actually decouple.

\subsec{Comparison with perturbative symmetry}

As we have seen above, the closed bosonic string possesses a
perturbative symmetry that contains the affine Kac-Moody algebra 
$\hat{su}(24)$. It is interesting to see how this algebra compares
with the conjecture that the non-perturbative bosonic string has a
$k_{27}$ symmetry. As we explained in section 2, the perturbative
subalgebra of $k_{27}$ is the very-extended algebra of $su(24)$. 

Our construction of the symmetry algebra $\hat{su}(24)$ was based on
the DDF-construction that is, by its nature, not Lorentz
invariant, but preserves only the $O(24)$ subgroup of the Lorentz
group that acts on the transverse coordinates. As such one should only
expect to see the part of the perturbative subalgebra of $k_{27}$ that
is  preserved by the $O(24)$ subgroup. In the Dynkin diagram of
figure~4, the $25$ nodes along the horizontal line of $k_{27}$
correspond to the Lorentz symmetries of $SO(1,25)$, and the space-time
directions used in the light-cone formalism are associated with the
two nodes that are on the far left. Thus we should only expect to see
the truncation of the perturbative subalgebra of $k_{27}$ where we
remove the $-2$ and $-1$ node. This is then precisely the affine
algebra $\hat{su}(24)$, in nice agreement with the analysis
above.

\newsec{Kac-Moody symmetries in the superstring}

The construction of the previous section generalises easily to the
case of the superstring, and we shall therefore be slightly sketchy
here. For simplicity we will restrict our attention to finding 
symmetries that act on the NS-NS sector of the theory only.

The physical state conditions in the NS-NS sector are given by 
$$L_0 \Psi = \bar{L}_0 \Psi = \half \Psi,
\eqno(4.1)$$
as well as  
$$
L_n \Psi = \bar{L}_n \Psi = 0 \qquad \rm{for\  all\  n>0}
\eqno(4.2)$$
and 
$$
G_r \Psi = \bar{G}_r \Psi = 0 \qquad \rm{for\  all\  r>0}.
\eqno(4.3)
$$
The physical states can be constructed in terms of the DDF operators
in a similar way to the closed bosonic string. We first enlarge the
space of states by doubling the zero modes, and then introduce the
chiral vertex operators in terms of which the DDF-operators are 
constructed. Since we are working in the NS-NS sector the zero modes
are the same as those of the bosonic theory and thus the construction
is completely analogous. In addition to the bosonic DDF-operators
$A^i_n$ and $\bar{A}^i_n$, there are now also fermionic operators
$B^i_r$ and $\bar{B}^i_r$ where $r\in \Zop+\half$ and $i=1,\ldots, 8$
denotes the transverse directions. All of these operators
(anti-)commute with $L_m$ and $G_r$, and therefore map states
satisfying equations  (4.1) -- (4.3) into themselves. However, as
before in the bosonic case, apart from the trivial operators  $A^i_0$
and $\bar{A}^i_0$, the single DDF operators do not leave the condition
$p_L=p_R$ invariant.   

The commutation relations among the different DDF operators are given
by   
$${\eqalign{
{} [ A^i_m, A^j_n ] & = m \, \delta^{ij}\, \delta_{m,-n}\,, \cr
{} [ A^i_m, B^j_r ] & = 0\,, \cr
{} \{B^i_r,B^j_s\} & = \delta^{ij}\, \delta_{r,-s} \,,}}
\eqno(4.4)$$
together with similar relations for the right-movers. Left- and
right-moving DDF operators again commute or anti-commute as
appropriate. 

Since the commutation relations for the bosonic DDF operators
are as before in the bosonic case, we can construct the generators
$R^{ij}_n$ that satisfy the commutation relations of the 
affine Kac-Moody algebra $\hat{su}(8) \oplus \hat{u}(1)$.  
The simplest fermionic DDF operators that preserve the momentum
matching condition are again bilinear in the fermionic DDF-operators,
and are given by 
$${\eqalign{
F^{ij}(r) & = B^i_{-r} B^j_{r} - \half \delta^{ij}\,, \cr
\bar{F}^{ij}(r) 
& = \bar{B}^{i}_{-r} \bar{B}^{j}_{r} - \half \delta^{ij} \,, \cr
G^{ij}(s) & = i B^i_{-s} \bar{B}^{j}_{-s} \,,}}
\eqno(4.5)$$
where the index $r$ of the first two sets of generators is assumed to
be positive, while $s$ is unrestricted. These generators are the 
fermionic analogues of $L^{ij}(n)$, $\bar{L} ^{ij}(n)$ and
$K^{ij}(n)$ of equations (3.20) and (3.21), respectively. 
Using the anti-commutation relations of equation (4.4) we find that
they obey the commutation relations  
$${\eqalign{
{}[F^{ij}(r),F^{kl}(s)] 
& = \delta_{r,s} 
   \left( \delta^{jk} F^{il}(r) - \delta^{il} F^{kj}(r)\right) \,,\cr
{}[F^{ij}(r), G^{kl}(s)]  & 
= \delta_{r,s} \delta^{j k} G^{il}(s) 
- \delta_{r,-s} \delta^{ik} G^{jl}(s)\,,}}
\eqno(4.6)$$
as well as 
$$
{\eqalign{
{}[\bar F^{ij}(r),\bar F^{kl}(s)] 
& = \delta_{r,s} 
\left( \delta^{jk} \bar F^{il}(r) 
     - \delta^{il} \bar F^{kj}(r)\right)\,,\cr 
{}[\bar F^{ij}(r), G^{kl}(s)]  & 
= \delta_{r,s} \delta^{jl} G^{ki}(s) 
- \delta_{r,-s} \delta^{il} G^{kj}(s)\,,}}
\eqno(4.7)$$
and
$$
{} [G^{ij}(r),G^{kl}(s)] = \delta_{r,-s} 
\left( \delta^{ik} \bar F^{jl}(r) + \delta^{jl} F^{ik}(r) \right)\,.
\eqno(4.8)
$$
Here $r>0$, while $s$ is positive in the first lines of (4.6) and 
(4.7), and unrestricted otherwise. These commutation relations are
identical to those of $L^{ij}(n)$, $\bar{L}^{ij}(n)$ and 
$K^{ij}(n)$. As such, these generators give rise, for each
positive $r$, to the Lie algebra of $su(16)\oplus u(1)$.  The
generators $F^{ij}(r)$ and $\bar{F}^{ij}(r)$ define the
algebra $u(8)\oplus u(8)$, and the generators $G^{ij}(s)$ with
$s=\pm r$ enhance  this algebra to $u(16)=su(16)\oplus u(1)$. 

In the type II superstring theories, the physical states are required
to be even under the two GSO-projections. While the $F^{ij}(r)$ and
$\bar{F}^{ij}(r)$ generators commute with the GSO-operators
$(-1)^F$ and $(-1)^{\bar{F}}$, the generator $G^{ij}$
anti-commutes with both of them. Thus in the GSO-projected type
IIA/IIB theory only the generators $F^{ij}(r)$ and 
$\bar{F}^{ij}(r)$ map physical states into one another. Thus
for each positive $r$, the relevant algebra is 
$u(8)\oplus u(8)$.  

As before, we could consider the generators $F^{ij}(r)$ and
$\bar{F}^{ij}(r)$ separately, but it is perhaps more natural 
to take the linear combination of generators that acts symmetrically
on left- and right-moving states. The corresponding generators form
the diagonal subalgebra, and are given as 
$${
S^{ij}(r) = F^{ij}(r) + \bar{F}^{ij}(r)\,.}
\eqno(4.9)$$
As for the closed bosonic string, we may combine the different copies
of the algebra $u(8)$ (for different values of $r$) into an affine 
Kac-Moody algebra by considering the infinite linear combination 
$$
S^{ij}_m = \sum_{r=\half}^{\infty} \zeta^{rm} \, S^{ij}(r) \,.
\eqno(4.10)$$
These modes then satisfy the commutation relations of
$\hat{su}(8)\oplus \hat{u}(1)$. 

The modes $R^{ij}_m$ and  $S^{ij}_m$ act separately on the states
generated by the bosonic and fermionic DDF-operators, respectively. It
is therefore natural to combine these modes as 
$$
U^{ij}_m = R^{ij}_m + S^{ij}_m \,.
\eqno(4.11)$$
The zero mode $U^{ij}_0$ then acts geometrically by construction, and
the complete set of modes defines the affine Kac-Moody algebra 
$\hat{su}(8)\oplus \hat{u}(1)$. Following the same argument as 
for the closed bosonic string, we expect this algebra to be a symmetry of
the perturbative string scattering amplitudes.

\subsec{Comparison to perturbative symmetries}

As before for the case of the bosonic string we can now compare the
symmetry we have constructed above with what one finds based on the
conjecture that M-theory is invariant under an $e_{11}$ symmetry. As
we have shown in section~2, the perturbative subalgebra is different
for the type IIA and IIB superstring theories. In the former case, the
perturbative sub-algebra is the very-extended $su(8)$ algebra. Taking
into account that the DDF construction only preserves the Lorentz
symmetries of the transverse directions, we should then expect to find
the subalgebra of the very-extended $su(8)$ algebra where we remove
the left most two dots of the Dynkin diagram in figure~2 (that are
associated to the light-cone directions). The resulting algebra is
then $\hat{su}(8)$, in nice agreement with the perturbative 
symmetry found above. 

For the IIB theory, the perturbative sub-algebra implied by
$e_{11}$ is very-extended $e_{7}$. Following the same procedure
as above, the part of this symmetry that is preserved by the
transverse Lorentz transformations is then $\hat{e_7}$. Since $e_7$
contains $su(8)$, it follows that $\hat{e_7}$ contains the affine
$\hat{su}(8)$ algebra. Hence, we find that the perturbative symmetry
found above in the NS-NS sector of the IIB string is a large part of
the symmetry that is predicted by the $e_{11}$ conjecture. 
\smallskip

Actually, on general grounds one may not necessarily expect to find
complete agreement between the  two symmetry algebras, even once one
has taken account of the non-Lorentz covariant nature of the DDF
construction. First of all, the arguments of \refs{\pwtwo} only
suggest that M-theory is invariant under $e_{11}$, but is quite
possible that this is in fact part of some larger symmetry based for
example on a Borcherds algebra. Secondly, the above perturbative
calculation only dealt with the NS-NS sector of the theory. It is
therefore possible that some of the symmetries found above may not
actually lift to the full theory; conversely there may be
additional symmetries that exchange states between the NS-NS and R-R
sectors. Given these considerations, the extent of the agreement
between the  perturbative symmetries  following from the $e_{11}$
conjecture, and those found in terms of the DDF construction is
substantial and gives support for the proposal that $e_{11}$ is indeed 
a symmetry of M-theory.

\newsec{Conclusions and discussions}

In this paper we have tested some predictions of the conjecture that
M-theory and the bosonic string possesses $e_{11}$ and $k_{27}$ as
symmetries, respectively. More specifically, we have determined the
perturbative subalgebras for Type IIA, IIB and the closed bosonic
string that are predicted by this proposal, and we have compared them
with the symmetry algebras that can be constructed in perturbative
string theory, using the DDF construction. Taking account of the
non-Lorentz covariance of the DDF-construction, we have found
remarkable agreement. 

At first sight it seems odd that the perturbative string theories
should have an (infinite dimensional) affine symmetry algebra, given
that there are only finitely many states at each mass level. However,
the affine algebras that appear in our construction are at level zero
(\ie\ have a trivial centre), and such algebras do possess non-trivial
finite-dimensional representations. 
\smallskip

The symmetry algebras that have been proposed for M-theory and the
closed bosonic string, $e_{11}$ and $k_{27}$, are rather special
Lorentzian algebras. In fact both are very-extended Lorentzian
Kac-Moody algebras whose finite dimensional semi-simple Lie algebras,
$e_{8}$ and $d_{24}$, are closely related to even self-dual Euclidean
lattices \refs{\gow}. As is well known even self-dual Euclidean
lattices only exist in dimensions $D=8n$, where $n=1,2,3,...$. The
first example therefore arises in dimension eight, where there is a
unique possibility, the root lattice of $e_8$. The corresponding
very-extended algebra is then $e_{11}$. In dimension $D=24$, on the
other hand, there are twenty-four such lattices, the so-called
Niemeier lattices \refs{\cs}. Apart from the Leech lattice that does
not have any vectors of length squared two, all other Niemeier
lattices are associated with finite dimensional semi-simple Lie
algebras. The root lattices of these Lie algebras are in general not
self-dual; however, they can be made self-dual by addition of a
specific set of conjugacy classes of the corresponding weight
lattice. Only two of the Lie algebras that arise in this context are
simple, namely $d_{24}$ and $a_{24}$. As we have seen above, the
very-extended algebra corresponding to the former Lie algebra is
$k_{27}$ that has been proposed as the symmetry algebra of the bosonic 
string.\footnote{$^\dagger$}{It might seem that the symmetries of the
closed bosonic string and M-theory arise in a different way. However,
the root lattice of $e_8$ can be obtained from the root lattice of
$d_8$ by addition of one of the spinor conjugacy classes of
$d_8$. From this point of view, the construction of the root lattice
of $e_8$ and the Niemeier lattice corresponding to $d_{24}$ is
completely analogous.} Thus, if we restrict our attention to
very-extended algebras that are associated to even self-dual lattices
in the way described above, we find almost uniquely the conjectured 
symmetry algebras of M-theory and the twenty-six dimensional bosonic
string. This suggests that even self-dual lattices play an important
r\^ole in the proper formulation of M-theory and the analogous theory
in twenty-six dimensions.

It is interesting to speculate about the significance of some of the
other symmetry algebras that can be constructed in a similar vein. If
instead of $d_{24}$ we considered $a_{24}$, the corresponding
very-extended Kac-Moody algebra would contain $a_{26}=su(27)$ as a 
subalgebra. If this very-extended Kac-Moody algebra was the symmetry
algebra of some theory, this theory would be expected to contain
gravity in twenty-seven dimensions. In fact, the corresponding theory
would have a low energy effective action that is probably just 
twenty-seven dimensional gravity, since gravity in $D$ dimensions
is believed to have the very-extended $a_D$ Kac-Moody algebra as a
symmetry \refs{\lw}.

\subsec{An $18$-dimensional theory}

Given the close relationship between the root lattices of $d_{8n}$ for 
$n=1,3$ and the conjectured symmetry algebras of M-theory and the
closed bosonic string, respectively, it is also interesting to
consider the case of the symmetry algebra that is associated to the root
lattice of $d_{16}$. As before, the corresponding even self-dual
lattice is obtained by adding to the root lattice of $d_{16}$ the
conjugacy class of weight vectors that contains one of the two spinor  
weights. (By the way, apart from the root lattice of $e_8\oplus e_8$
this is the only even self-dual lattice in $16$ dimensions.) 
Let us denote the corresponding very-extended Kac-Moody algebra
by $k_{19}$. Its Dynkin diagram is of the same form as that of
figure~4, except that there are only $17$ nodes along the horizontal. 

Given the $k_{19}$ symmetry we can deduce the low energy effective
action of the proposed new theory, using similar arguments to those in
\refs{\pwtwo}. The nodes along the horizontal line
are just those of $a_{17}$ and therefore correspond to the presence of
a graviton. The vertical node on the right corresponds to a rank two
anti-symmetric tensor of $a_{17}$  which we identify with a gauge
field, $B_{a_1a_2}$, while the vertical node to the left corresponds
to a rank fourteen anti-symmetric tensor that we denote by
$B_{a_1\ldots a_{14}}$; it can be identified with the dual gauge field
of the anti-symmetric tensor $B_{a_1a_2}$. 

By construction the algebra $k_{19}$ has rank nineteen. The graviton
is associated to the group GL(18), and its associated generators
provide the Cartan subalgebra of $a_{17}$, as well as one additional
Cartan generator $D$ of $k_{19}$. The remaining Cartan generator,
which we denote by $R$, corresponds then to an additional scalar
field, denoted by $\phi$. The low energy effective action of the  
proposed new theory in eighteen dimensions therefore contains  
gravity, a second rank tensor gauge field $B_{a_1a_2}$, its dual gauge
field $B_{a_1\ldots a_{14}}$, a scalar $\phi$ and its dual 
$\phi_{a_1\ldots a_{16}}$. The field content is therefore very
analogous to that of the closed bosonic string in twenty-six
dimensions.  

A theory with just such a field content was constructed as a
non-linear realisation in reference \refs{\pwtwo}. The relevant Lie
algebra is generated by $K^a{}_b$, $R$, $R^{a_1a_2}$,
$R^{a_1\ldots a_{14}}$ and  $R^{a_1\ldots a_{16}}$, with
commutation relations 
$$\eqalign{
[K^a{}_b, R^{a_1\ldots a_p}] & = \delta^{a_1}{}_b
R^{b a_2\ldots a_p}+\cdots\,,\cr
[R, R^{a_1\ldots a_p}] & = c_p \, R^{a_1\ldots a_p}\,, \cr
[R^{a_1a_2}, R^{a_1\ldots a_{14}}] & = 2\, R^{a_1\ldots a_{16}} \,,}
\eqno(5.1)$$
where $c_2={1\over 4}=-c_{14},\ c_0=0=c_{16}$, and all other
commutators vanish. The corresponding effective action, after the
elimination of the dual fields from the equations of motion, is then
given by 
$$ 
S = \int d^{18}x \,\left(R 
-{1\over2}\, \partial_\mu\phi\, \partial^\mu \phi
-{1\over 12}\, e^{\beta\phi}\, H_{\mu\nu\lambda}\, 
H^{\mu\nu\lambda}\right)\,,
\eqno(5.2)$$
where $\beta={1\over \sqrt 2}$. 

It is encouraging that upon dimensional reduction of this effective
action on a torus some part of the underlying Kac-Moody symmetry
becomes manifest. Indeed, it was shown in \refs{\lw} that this is the
case provided that $\beta$ is chosen as above. Although the class of 
actions that give rise to Kac-Moody symmetries upon reduction is very 
restricted, the generalisation of the above action to $D$ dimensions
also leads to a Kac-Moody symmetry provided one chooses 
$\beta = \sqrt {{8\over (D-2)}}$. It is an amusing coincidence that 
for $D=8n+2$ this expression simplifies to $\beta=\sqrt {{1\over n}}$.
Furthermore, for $n=1, 2, 3$ the above action describes the (gravity,
scalar and anti-symmetric 2-form sector) of the low energy effective
action for the ten dimensional type II superstring, the eighteen
dimensional string theory proposed here, and the twenty-six
dimensional closed bosonic string theory, respectively.

Next we want to show that the Kac-Moody algebra that is associated with
this theory is in fact $k_{19}$. Our arguments will be similar to
those that were used to deduce that $e_{11}$ and $k_{27}$ are the
symmetries for the maximal supergravity theory in eleven dimension and
the closed bosonic string theory in twenty-six dimensions,
respectively. If the theory is to admit a Kac-Moody algebra as a
symmetry then the generators  
$$K^a{}_b\,,\quad a<b\,, \qquad
R^{a_1a_2}\,,\qquad
R^{a_1\ldots a_{14}}\,, \qquad
R^{a_1\ldots a_{16}}
\eqno(5.3)$$
must be some of its positive roots. Furthermore, we can expect its Cartan
sub-algebra to contain 
$$H_a=K^a{}_a-K^{a+1}{}_{a+1}\,,\quad a=1, \ldots, 17\,, 
\qquad D=\sum_{a=1}^{18} K^a{}_a\,, \quad R\,.
\eqno(5.4)$$
The set of equation (5.3) is generated by the multiple commutators of 
$$E_a=K^a{}_{a+1}\,,\quad a=1\ldots 17\,, \qquad
E_{18}=R^{17 18}\,, \qquad
E_{19}=R^{5\ldots 18}\,,
\eqno(5.5)$$
which, as the notation suggests, we should identify with the simple
roots of the proposed Kac-Moody algebra. Next we want to find  Cartan
generators (that can be expressed in terms of the generators of
equation (5.4)), which lead to an acceptable  Cartan matrix for a
Kac-Moody algebra and contain the subalgebra $a_{17}$ in the
appropriate way. One can readily show that the choice 
$$\eqalign{
H_a & = K^{a}{}_a - K^{a+1}{}_{a+1} \,, \qquad  a=1,\ldots 17\,,\cr
H_{18} & = K^{17}{}_{17}+K^{18}{}_{18} - 
{1\over 8} D+R\,,\cr
H_{19} & =K^{5}{}_{5}+\ldots+K^{18}{}_{18} - 
{7\over 8}D-R}
\eqno(5.6)$$
satisfies these requirements. It follows from these equations and 
the Kac-Moody algebra relation $[H_a,E_b]=A_{ab} E_b$
that the Cartan matrix $A_{ab}$ is that for the very-extended $d_{16}$ 
Kac-Moody algebra, \ie\ $k_{19}$.
\smallskip

Thus we have found convincing evidence that, should there exists a
theory in eighteen dimensions that is associated with the
very-extended $d_{16}$ algebra, then this theory would have a low
energy effective action that is given by equation (5.2). A graviton
and second rank tensor gauge field in D dimensions have 
$\half (D-1)(D-2) -1$ and $\half (D-2)(D-3)$ degrees of freedom,
respectively. Together with 
the single scalar field we therefore find that the number of bosonic
degrees of freedom is $256$. If the theory were supersymmetric we
would require a matching number of fermionic degrees of freedom. One
of these fermions would have to be the gravitino which has
$2^{{(D-2)\over 2}}(D-3)\ r$ degrees of freedom, where $r={1\over 2}$ or
$r=1$ if it is a Majorana-Weyl or Majorana spinor, respectively. For
$D=18$ we find that the gravitino has ${15\over 2} . 256$ degrees of
freedom even if it were a Majorana-Weyl spinor, and we may conclude
that this theory could not possess space-time supersymmetry.  

Given the presence of the rank two tensor gauge field and the graviton
it would seem reasonable to suppose that this theory was a type of
string theory. Indeed the count of states is just that for a
closed bosonic string theory. Clearly, this theory would posses an 
anomaly and one would have to  add  a conformal field theory which did
not contribute states in the massless sector and lead to a modular
invariant partition function. Some time ago an eighteen dimensional
string theory has been suggested \refs{\ens} within the context of
the mechanism proposed for recovering supersymmetric string theories
from the  closed bosonic string in twenty-six dimensions. However, the
theory of \refs{\ens} has only $136$ massless bosons, and therefore
does not seem to be the same as that considered here.

\subsec{Relation to Borcherds algebras of even self-dual Lorentzian
lattices} 

Finally, let us comment on the relation of the symmetry algebras we
have found to Borcherds algebras \refs{\borcherds}. Recall that to
every even self-dual Euclidean lattice $\Lambda$ of dimension $D$, one
can naturally associate an even self-dual Lorentzian lattice of
dimension $D+2$. This lattice is simply $\Lambda\oplus \II^{1,1}$,
where $\II^{1,1}$ denotes the even self-dual Lorentzian lattice in
$1+1$ dimensions. Even self-dual Lorentzian lattices exist in
dimensions $D=8n+2$, and for each $n$ there is a unique such lattice
that is usually denoted by $\II^{1,8n+1}$. It is easy to see that if 
$\Lambda$ is the  root lattice of a semi-simple Lie algebra, then
$\Lambda\oplus\II^{1,1}$ is the root lattice of the over-extension of
this algebra \refs{\go}. As we have seen above, the symmetry algebras
$e_{11}$ and $k_{27}$ originate from finite dimensional semi-simple
Lie algebras that are associated with even self-dual Euclidean
lattices. The corresponding over-extensions are therefore related to
even self-dual Lorentzian lattices, and the very-extended algebras
that are of relevance here can be obtained from these by adding one
additional node to the Dynkin diagram.   

For the case of the symmetries of M-theory, the relevant over-extended
algebra is the hyperbolic algebra $e_{10}$, whose root lattice is the
unique even self-dual Lorentzian lattice $\II^{1,9}$. It has been
known for some time that the hyperbolic Lie algebra $e_{10}$ is
actually rather complicated --- for example its root multiplicities
are only known at low levels. On the other hand, the corresponding
Borcherds algebra, \ie\ the algebra of physical states of the bosonic
string compactified on the lattice $\II^{1,9}$, has known root
multiplicities and contains $e_{10}$ \refs{\eten,\etenp}. One
may therefore guess that the actual symmetry algebra of M-theory may
indeed contain this Borcherds algebra rather than just $e_{10}$. 

Similarly, for the case of the closed bosonic string, the
over-extension of $d_{24}$ is a $26$-dimensional Lorentzian Kac-Moody
algebra $k_{26}$. Its root lattice is the sublattice
$\Lambda_{d_{24}}\oplus\II^{1,1}$ of the unique even
self-dual Lorentzian lattice $\II^{1,25}$. From the point of view of
twenty-four dimensions, the corresponding Niemeier lattice
$\Lambda^N_{d_{24}}$ can be obtained from the root lattice of 
$d_{24}$ by adding the conjugacy class of the $d_{24}$ 
weight space that contains one of the spinors. Since the shortest
vector in this conjugacy class  has length squared six, it is
conceivable that these vectors are not visible in the low-energy 
analysis that underlies \refs{\pwtwo}, and that the symmetry algebra
should actually contain all the roots of $\II^{1,25}$. Furthermore,
this would then suggest that the actual symmetry algebra contains the
corresponding Borcherds algebra, which, in this case, is just the fake
Monster algebra. The fake monster algebra was suggested as a symmetry of
the closed  bosonic string compactified on the torus $\II^{1,25}$ 
\refs{\m,\pwthree}. The philosophy of \refs{\pwtwo} is that symmetries
that are seen upon dimensional reduction are actually symmetries of
the underlying unreduced theory. It is an encouraging sign that 
the symmetry algebra $k_{27}$ that was found from the low energy
effective action in twenty-six dimensions, contains the algebra
$k_{26}$ which, as we have just explained, is closely related to the
fake Monster symmetry that emerges upon dimensional reduction. 

In the above we have suggested that we should replace the
over-extended subalgebra of the symmetry algebra by the Borcherds
algebra associated to the corresponding even self-dual (Lorentzian)
lattice. This then also implies that the full symmetry algebra is not 
just a very-extended Kac-Moody algebra, but actually some extension
of a Borcherds algebra. For the case of the superstring it is
conceivable that the relevant algebra is just the usual Borcherds
algebra based on the root lattice of $e_{11}$, but this is not
possible for $k_{27}$ since the dimension of the latter lattice is
$27$ and thus the no-ghost theorem of string theory (that plays a
crucial role for Borcherds' construction) fails. At any rate, it seems
unlikely that the full symmetry algebra is just a Borcherds algebra,
given that the underlying theories in $11$ and $27$ dimensions are
probably theories of membranes rather than strings. Thus we expect
that the very-extended symmetry algebras should be replaced by some
generalisation of Borcherds algebras whose vertex operators are based on
membranes. It would be very interesting to discover what class of
algebras membrane vertex operators lead to.  

The idea that the underlying algebras are some sort of Borcherds
algebras is also attractive from a more conceptual point of view. As
is well known, the root lattices of inequivalent Kac-Moody algebras
can be the same (for some explicit examples see for example
\refs{\gow}), and thus the root lattice does not uniquely define the
corresponding Kac-Moody algebra. Rather, in order to define the
Kac-Moody algebra, one must also specify a preferred set of basis
vectors, namely the simple roots whose scalar products define the
Cartan matrix of the Kac-Moody algebra. On the other hand, the
Borcherds algebras can be uniquely constructed from their root
lattices (they are just the algebra of physical states of the
string compactification on the corresponding lattice), and their root
multiplicities are known (since they are just the number of physical
states of a given momentum). It would be very interesting to find more
evidence for these speculations.

\appendix{A}{The simple roots of $e_8$}

The root lattice of $e_8$ is an eight-dimensional Euclidean even
self-dual lattice. It is spanned by the simple roots which can be
taken to be given by the following vectors in $\Rop^8$.
$$
\eqalign{
\alpha_1 & = (0,0,0,0,0,1,-1,0)\,, \cr
\alpha_2 & = (0,0,0,0,1,-1,0,0)\,, \cr
\alpha_3 & = (0,0,0,1,-1,0,0,0)\,, \cr
\alpha_4 & = (0,0,1,-1,0,0,0,0)\,, \cr
\alpha_5 & = (0,1,-1,0,0,0,0,0)\,, \cr
\alpha_6 & = (-1,-1,0,0,0,0,0,0)\,, \cr
\alpha_7 & = \left({1\over 2},{1\over 2},{1\over 2},{1\over 2},
{1\over 2},{1\over 2},{1\over 2},{1\over 2}\right)\,, \cr
\alpha_8 & = (1,-1,0,0,0,0,0,0)\,. \cr}
\eqno(\hbox{A.1})$$

\footatend\vfill\supereject\immediate\closeout\rfile\writestoppt
\baselineskip=14pt\centerline{{\bf References}}\bigskip{\frenchspacing%
\parindent=20pt\escapechar=` \input refs.tmp\vfill\eject}\nonfrenchspacing

\bye